\RequirePackage{amsmath}

\documentclass[runningheads]{llncs}

\usepackage[]{algorithm2e}
\usepackage[noend]{algpseudocode}
\usepackage{amssymb}
\usepackage[toc,page]{appendix}
\usepackage{cite}
\usepackage{commands-tam}
\usepackage{float}
\usepackage{graphicx}
\usepackage[utf8]{inputenc}
\usepackage{multirow}
\usepackage{wrapfig}
\usepackage{subcaption} 
\usepackage[textsize=scriptsize]{todonotes}

\usepackage{url}

\newcommand{\Gtri}{\ensuremath{G_{\Delta}}}
\newcommand{\eflag}{\ensuremath{\epsilon}}
\newcommand{\idle}{\ensuremath{\textsf{idle}}}
\newcommand{\expand}{\ensuremath{\textsf{expand}}}
\newcommand{\contract}{\ensuremath{\textsf{contract}}}
\newcommand{\handover}{\ensuremath{\textsf{handover}}}
    
\title{Simulation of Programmable Matter Systems Using Active Tile-Based Self-Assembly
\thanks{Daymude and Richa are funded in part by the National Science Foundation under awards CCF-1422603, CCF-1637393, and CCF-1733680. Alumbaugh and Patitz are funded in part by National Science Foundation Grants CCF-1422152 and CAREER-1553166.}}
\titlerunning{Simulating Programmable Matter with Self-Assembly}

\author{John Calvin Alumbaugh\inst{1} \and
Joshua J. Daymude\inst{2}\orcidID{0000-0001-7294-5626} \and
Erik D. Demaine\inst{3} \and
Matthew J. Patitz\inst{1}\orcidID{0000-0001-9287-4028} \and
Andr\'ea W. Richa\inst{2}}
\authorrunning{Alumbaugh, Daymude, Demaine, Patitz, and Richa}

\institute{Department of Computer Science and Computer Engineering, University of Arkansas; \email{mpatitz@self-assembly.net} \and
Computer Science, CIDSE, Arizona State University; \email{\{jdaymude,aricha\}@asu.edu} \and
MIT Computer Science and Artificial Intelligence Laboratory; \email{edemaine@mit.edu}}

\begin{document}

\maketitle

\begin{abstract}
Self-assembly refers to the process by which small, simple components mix and combine to form complex structures using only local interactions. Designed as a hybrid between tile assembly models and cellular automata, the \emph{Tile Automata (TA) model} was recently introduced as a platform to help study connections between various models of self-assembly. However, in this paper we present a result in which we use TA to simulate arbitrary systems within the \emph{amoebot model}, a theoretical model of programmable matter in which the individual components are relatively simple state machines that are able to sense the states of their neighbors and to move via series of expansions and contractions.
We show that for every amoebot system, there is a TA system capable of simulating the local information transmission built into amoebot particles, and that the TA ``macrotiles'' used to simulate its particles are capable of simulating movement (via attachment and detachment operations) while maintaining the necessary properties of amoebot particle systems. The TA systems are able to utilize only the local interactions of state changes and binding and unbinding along tile edges, but are able to fully simulate the dynamics of these programmable matter systems.
\end{abstract}

\keywords{programmable matter \and simulation \and self-assembly \and tile automata \and amoebot model}

\section{Introduction} \label{sec:intro}

Theoretical models of self-assembling systems are mathematical models that allow for the exploration of the limits of bottom-up construction and self-assembly via simple (usually square) tiles.
There are a wide variety of tile-based models of self-assembly (e.g.,~\cite{Winf98,KaoSchweller08,jSignals,jDuples,SingleNegative,Polygons,Polyominoes}), each with differing constraints and dynamics, resulting in great variations in the relative powers between systems.
One of the easiest ways to evaluate their relationships is to use notions of simulation to attempt to simulate one model by another, and this has led to the creation of a ``complexity hierarchy'' of self-assembly models and categories of systems~\cite{WoodsIUSurvey,IUSA,2HAMIU,Signals3D,IUNeedsCoop}.

Another category of theoretical models attempts to capture the dynamics of so-called \emph{programmable matter}, in which small and simple, but dynamic and mobile, components are able to interact with each other to form structures, perform tasks and computations, etc. \cite{Nubots,Daymude2019}.

This paper attempts to bridge the divide between these categories of models, showing how self-assembling tiles can mimic the behaviors of programmable matter. Specifically, we demonstrate how the recently introduced \emph{Tile Automata (TA) model}~\cite{freezing} can be used to simulate the \emph{amoebot model}~\cite{Daymude2019}. In the TA model, the fundamental components are unit square tiles which form structures by attaching and forming bonds, and can also change states based on their own states and those of their neighbors, causing them to be able to form new bonds or to remove existing bonds. The basic components in the amoebot model are \emph{particles} which can also change their states based on the current states of themselves and their neighbors, but which can also move via series of expansions and contractions. While the components of both models rely only upon local information and communication, the design goals of their systems tend to differ fundamentally. The main goal of TA systems is to self-assemble into target structures, but amoebot systems have been used to solve system-level problems of movement and coordination (e.g., shape formation~\cite{Derakhshandeh2016}, object coating~\cite{Derakhshandeh2017}, leader election~\cite{Daymude2017}, gathering~\cite{Cannon2016}, bridging gaps~\cite{Arroyo2018}, etc.). We present a construction in which constant-sized assemblies of TA tiles, called \emph{macrotiles}, assemble and disassemble following the rules of the TA model and are able to simulate the behaviors of individual amoebot particles. Via carefully designed processes of building and breaking apart assemblies, they are collectively able to correctly simulate the full dynamics of amoebot systems. We thus show how the dynamics of systems of self-assembling tiles with the ability to form and break bonds can be harnessed to faithfully simulate the dynamics of collections of programmable matter particles capable of local communication and motion. Not only does this provide a way to connect and leverage existing results across models, this also provides a new paradigm for designing systems to accomplish the goals of programmable matter. It additionally allows amoebots to serve as a higher-level abstraction for designing systems exhibiting complex behaviors of programmable matter but with a translation to implementation in TA.

The paper is organized as follows.
Section~\ref{sec:TAmodel} presents a high-level definition of the TA model, and Section~\ref{sec:amoebotmodel} provides a full mathematical definition for the amoebot model. (We note that this is the first full mathematical definition for the amoebot model and thus is also a contribution of this paper.)
Section~\ref{sec:simoverview} gives the formal definition, preliminaries, and overview of the simulation of the amoebot model by TA, while Section~\ref{sec:simmove} gives more of its details. A brief discussion and conclusion are given in Section~\ref{sec:conclude}, and a Technical Appendix contains a more rigorous definition of the TA model, as well as low-level technical details about the construction.

\vspace{-5pt}
\section{The Tile Automata Model} \label{sec:TAmodel}
\vspace{-5pt}

The Tile Automata model seeks to connect and evaluate the differences between some of the seemingly disparate models of tile-based self assembly by combining components of the \emph{Two Handed Assembly Model} (2HAM) of self-assembly with a rule set of local state changes that are similar to asynchronous cellular automata. This section provides an overview of the TA model which is sufficient for the purposes of this paper, however a more thorough and detailed definition of the TA model is available in the technical appendix, which is based on \cite{freezing}.

\begin{figure}
    \centering
    \vspace{-5pt}
    \includegraphics[width=4.5in]{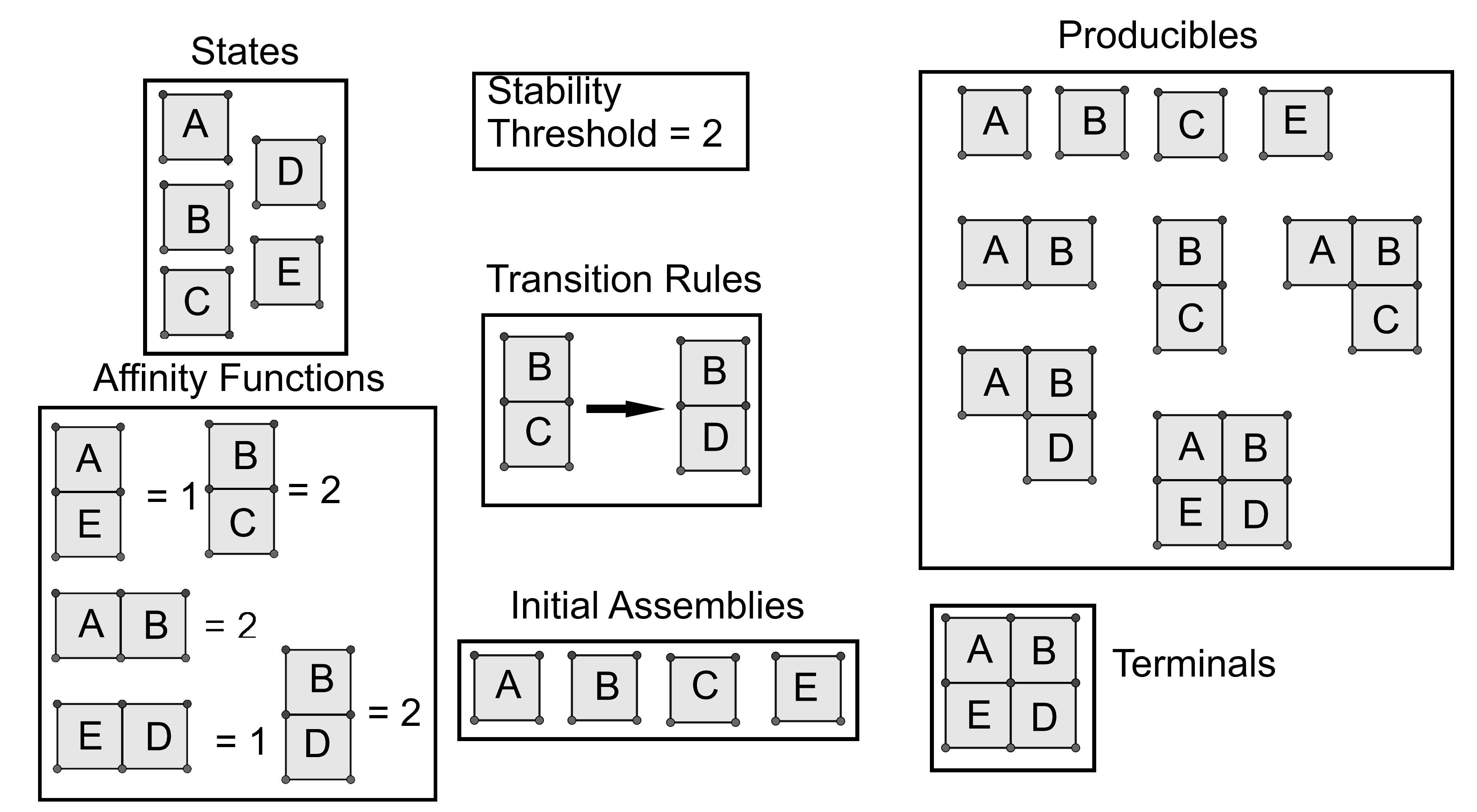}
    \caption{Example of a TA system. The five components that define a TA system constitute the left and middle columns of this figure, while the rightmost boxes indicate producible and terminal assemblies.}
    \label{fig:TA_ex}
    \vspace{-5pt}
\end{figure}

The Tile Automata Model has many similarities with other tile based self assembly systems. Tiles, the fundamental units of this model that interact with one another, use only local information, in this case the \textit{state} of their neighbors. Tiles exist as a stateful unit square centered on a point on the square lattice over the integers in two dimensions, so that a tile's coordinates $(x, y) \in \mathbb{Z}^2$. Tiles may form bonds with adjacent neighbors via attaching to one another according to the \textit{affinity function}, which defines a set of two states and either a vertical or horizontal relative orientation (denoted as $\perp$ and $\vdash$, respectively) as well as an attachment strength. A connection between tiles or groups of connected tiles must have the property of $\tau$ stability to persist. Every TA system has defined an integer \textit{stability threshold} or $\tau$ that represents the minimum strength bond with which tiles must be bound in order to be $\tau$ stable. Two adjacent tiles of states $s, s'$, with the tile of state $s$ directly to the right of the tile of state $s'$, will form an attachment if there exists a rule in the affinity function $(s' \vdash s \geq \tau)$. An \textit{assembly} is a $\tau$ stable connected set of TA tiles, with the property that there exists no way to separate the tiles without breaking bonds of at least $\tau$ strength. Further, a pair of tiles may \textit{transition} according to a transition rule that takes as input two adjacent tiles (oriented by either $\perp$ or $\vdash$) and outputs new states for those tiles. So the tiles in our example of $s' \vdash s$ may transition to states $t \vdash s$ if there exists a rule in the set of transition rules provided in the definition of a TA system of the form $(s', s, t, s, \vdash)$, where $(s', s)$ are the input states, $(t, s)$ are the output states, and $\vdash$ is their relative orientation.

\begin{figure}
    \begin{center}
    \vspace{-5pt}
    \includegraphics[width=3.0in]{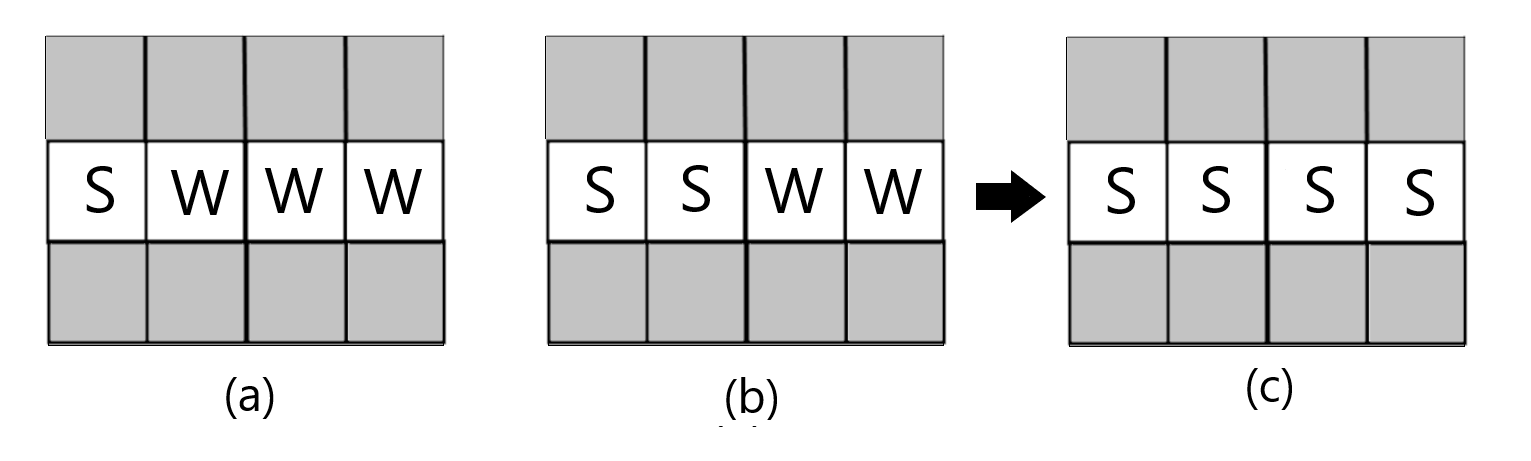}
    \caption{Depiction of signal \emph{S} being passed down a wire. The W tiles represent wires in their default state, and the grey tiles above and below the wire are filler tiles. Starting in state in (a) and a transition rule (SW $\vdash$ SS), the signal propagates down the wire in (b) and (c).}
    \label{fig:wire_transmission}
    \end{center}
\vspace{-5pt}
\end{figure}

\vspace{-5pt}
\subsection{Wire Transmission}
One of the most useful aspects of the Tile Automata model is the tiles' ability to transition states based on local information. This capability makes communication from one group of tiles to another, non-adjacent group easy, with a structure we will call a \emph{wire}. A wire in TA is a contiguous line of tiles from one group of tiles to another, usually surrounded by inert filler tiles so as to avoid interference with the signal being transmitted. (See Figure~\ref{fig:wire_transmission} for an example.)

\vspace{-5pt}
\section{The Amoebot Model} \label{sec:amoebotmodel}

Introduced in \cite{amoebots}, the amoebot model is an abstract computational model of \emph{programmable matter}, a substance that can change its physical properties based on user input or stimuli from its environment.
The amoebot model envisions programmable matter as a collection of individual, homogeneous computational elements called \emph{particles}.
In what follows, we extend the exposition of the model in~\cite{Daymude2019} to the level of formality needed for our simulation.



\begin{figure}[tbh]
\centering
\begin{subfigure}{.33\textwidth}
	\centering
	\includegraphics{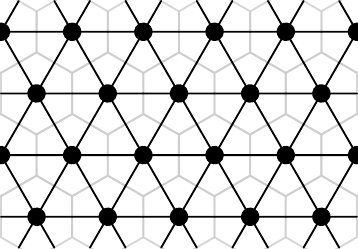}
	\caption{}
	\label{fig:amoebotlattice}
\end{subfigure}%
\begin{subfigure}{.33\textwidth}
	\centering
	\includegraphics{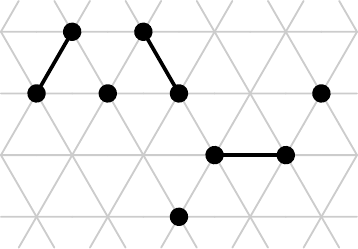}
	\caption{}
	\label{fig:amoebotparticles}
\end{subfigure}%
\begin{subfigure}{.33\textwidth}
	\centering
	\includegraphics{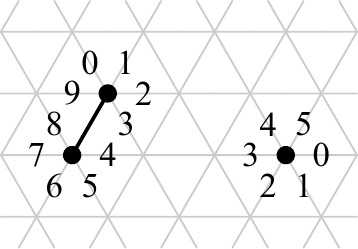}
	\caption{}
	\label{fig:amoebotlabels}
\end{subfigure}
\caption{(a) A section of the triangular lattice $\Gtri$ (black) and its dual, the hexagonal tiling (gray). (b) Expanded and contracted particles (black dots) on $\Gtri$ (gray lattice). Particles with a black line between their nodes are expanded. (c) Two particles with different orientations. The expanded particle's tail port would be 6 if its head were the upper node; the contracted particle's tail port is $\varepsilon$.}
\vspace{-15pt}
\label{fig:amoebotmodel}
\end{figure}

Any structure a particle system can form is represented as a subgraph of an infinite, undirected graph $G = (V, E)$ where $V$ is the set of positions a particle can occupy and $E$ is the set of all atomic movements a particle can make.
Each node in $V$ can be occupied by at most one particle at a time.
This work further assumes the \emph{geometric amoebot model} where $G = \Gtri$, the triangular lattice with nearest neighbor connectivity (see Fig.~\ref{fig:amoebotlattice}). This lattice is preferred for work in the 2D plane, as it allows for a maximum of nearest neighbor connectivity for particles moving step wise around the perimeter of the particle swarm. Particles attempting to move around a ``corner" of a particle swarm risk disconnection with the neighborhood implied by nearest neighbor connectivity on the square lattice.
Each particle occupies either a single node in $V$ (i.e., it is \emph{contracted}) or a pair of adjacent nodes in $V$ (i.e., it is \emph{expanded}), as in Fig.~\ref{fig:amoebotparticles}.
Two particles occupying adjacent nodes of $\Gtri$ are \emph{neighbors}. We further will define a group of particles as a \emph{particle system}.

Each particle keeps a collection of \emph{ports} --- one for each edge incident to the node(s) it occupies --- that have unique labels from its own perspective.
Contracted particles have six ports while expanded particles have ten (see Fig.~\ref{fig:amoebotlabels}).
The particles are assumed to have a common sense of clockwise direction (a.k.a. chirality), but do not share a coordinate system or global compass.
Thus, particles can label their ports in clockwise order starting from a local direction 0, but may have different \emph{orientations} in $O = \{0,1, ..., 5\}$ encoding their offsets for local direction 0 from global direction 0 (to the right).

For example, in Fig.~\ref{fig:amoebotlabels}, the particle on the right has orientation 0 (i.e., it agrees with the global compass) while the particle on the left has orientation 4 (i.e., its local direction 0 is global direction 4).
When a particle expands, it keeps its port labeling consistent by assigning label 0 to a port facing local direction 0 and then labeling the remaining ports in clockwise order.\footnote{Note that there may be ambiguity in choosing a port facing local direction 0; e.g., in Fig.~\ref{fig:amoebotlabels}, both port 0 and port 8 face local direction 0.
In this case, the port facing local direction 0 and ``away'' from the particle is labeled 0.}
In this way, it can recover its original labeling when it later contracts.
A particle $p$ communicates with a neighbor $q$ by placing a flag from the constant-size alphabet $\Sigma$ on its port facing $q$.
This can be thought of as $p$ sending a message for $q$ to read when $q$ is next activated.
Conversely, $p$ receives information from $q$ by reading the flag $q$ has placed on its port facing $p$.
The flag alphabet $\Sigma$ is assumed to contain the ``empty flag'' $\eflag$ to be used when no information is being communicated.

Particles move via a series of \emph{expansions} and \emph{contractions}: a contracted particle can expand into an unoccupied adjacent node to become expanded, and may then contract to occupy a single node once again.
An expanded particle's \emph{head} is the node it last expanded into and the other node it occupies is its \emph{tail}; a contracted particle's head and tail are the same.
If an expanded particle contracts into its head node, it has moved.
Otherwise, contracting back into its tail node can be thought of as the particle exploring a potential location to which it could expand but deciding not to over the course of two activations.
Neighboring particles can coordinate their movements in a \emph{handover}, which can occur one of two ways.
A contracted particle $p$ can ``push'' an expanded neighbor $q$ by expanding into one of the nodes occupied by $q$, forcing $q$ to contract.
Alternatively, an expanded particle $q$ can ``pull'' a contracted neighbor $p$ by contracting, forcing $p$ to expand into the node it is vacating.
During its movements, each particle maintains a \emph{tail port} in $T = \{0,1,\dots,9\}\ \cup \{\varepsilon\}$ denoting the port furthest from its head if it is expanded or $\varepsilon$ if it is contracted (see Fig.~\ref{fig:amoebotlabels}).
This information serves as the particle's memory about whether or not it is expanded, and, if so, what direction its tail is relative to its head.

More formally, the set of all possible movements is $M = \{\idle\} \cup \{\expand_i : i \in {0,1,\dots,5}\} \cup \{\contract_i : i \in {0,1,...,9}\} \cup \{\handover_i : i \in {0,1,...,5}\}$.
An $\idle$ move simply means the particle does not move.
If a particle $p$ performs $\expand_i$, $p$ expands into the node its $i$-th port faces only if $p$ is contracted and that node is unoccupied.
If a particle $p$ performs $\contract_i$, $p$ contracts out of the node incident to its $i$-th port only if $p$ is expanded.
The $\handover_i$ moves are not push or pull handover specific, nor do they actually perform the handover movements described above.
Instead, a particle $p$ performs $\handover_i$ when it initiates a handover with the neighbor its $i$-th port faces, say $q$.
This initiation only succeeds if a neighboring particle $q$ actually exists and $p$ is contracted while $q$ is expanded (or vice versa).
To aid in executing the initiated handover --- which will be described shortly --- each particle keeps an \emph{expansion direction} in $E = {0,1,...,5} \cup \{\epsilon\}$ denoting the local direction it would like to expand in or $\epsilon$ if no expansion is needed.

The amoebot model assumes that particle systems progress by individual particles performing atomic actions asynchronously, where each particle independently and continuously executes its own instance of the given algorithm at potentially varying speeds.
Assuming any conflicts that may arise in this concurrent execution are resolved --- as is the case in the amoebot model, see~\cite{Daymude2019} --- a classical result under the asynchronous model states that there is a sequential ordering of atomic actions producing the same end result.
Thus, we assume there is an \emph{activation scheduler} responsible for activating exactly one particle at a time.
This scheduler is assumed to be \emph{fair}: each particle is assumed to be activated infinitely often.
When a particle $p$ is activated by the scheduler, it computes its \emph{transition function} $\delta$ and applies the results:

\[\delta : Q \times \Sigma^{10} \times T \times E \to \mathcal{P}(Q \times \Sigma^{10} \times T \times E \times M).\]

For a given algorithm under the amoebot model, $Q$ is a constant-size set of particle \emph{states} while the flag alphabet $\Sigma$, the tail ports $T$, the expansion directions $E$, and the movements $M$ are as defined above.
The transition function $\delta$ allows a particle to use its state, its neighbors' flags facing it, its tail port, and its expansion direction to update these values and decide whether to move. If $\delta$ maps a unique input to multiple outputs, one output set is chosen arbitrarily.
$\delta$ largely depends on the algorithm being executed; here, we describe a few general rules for $\delta$ our simulation will consider.
Suppose that $\delta(q, (f_0, f_1, \ldots, f_9), t, e) = (q', (f_0', f_1', \ldots, f_9'), t', e', m')$.
\begin{itemize}
    \item The movement $m'$ must be valid according to the defined movement rules; e.g., if $m' = \expand_i$, particle $p$ must be contracted and the node its $i$-th port faces must be unoccupied.
    \item If $t = \varepsilon$, then $f_6 = \cdots = f_9 = \eflag$; i.e., if particle $p$ is contracted, it cannot set flags for ports it doesn't have. This holds also for $t'$ and $(f_6', \ldots, f_9')$.
    \item If $e \neq \epsilon$, then $t = \varepsilon$; i.e., particle $p$ can only intend to expand in local direction $e$ if it is contracted. This holds also for $e'$ and $t'$.
    \item If $m' = \expand_i$, then $t' \neq \varepsilon$ and $e' = \epsilon$; i.e., if particle $p$ expands in local direction $i$, it will be expanded (setting $t'$ to the label opposite $i$ after expansion) and should not intend to expand again immediately.
    \item If $e \neq \epsilon$, then either $m' = \expand_e$ or $m' = \idle$. That is, if particle $p$ intends to expand in local direction $e$ this activation, it either does so or has to wait.
    \item If $m' = \contract_i$, then $t' = \varepsilon$.
\end{itemize}

It remains to describe how handovers are executed with respect to $\delta$.
In a concurrent execution, a handover is performed as a coordinated, simultaneous expansion and contraction of two neighboring particles.
In our sequential setting, however, we instead use a local synchronization mechanism to ensure the contracting particle moves first, followed by the expanding particle.
To achieve this, we make one change to the scheduler.
Whenever a particle $p$ returns a movement $m' = \handover_i$ as output from $\delta$, the scheduler finds the neighbor $q$ facing the $i$-th port of $p$ and ensures that the next three particles to be activated are $q$, then $p$, then $q$ again.\footnote{Note that this forced scheduling is simply a result of our formalism and does not alter or subvert the underlying asynchrony assumed by the amoebot model.}
We first describe a pull handover initiated by an expanded particle $p$ with a contracted neighbor $q$.
\begin{enumerate}
    \item Suppose $p$ is chosen by the scheduler.
    Based on its state, its neighbors' flags facing it, its tail port indicating it is expanded, and its (empty) expansion direction, suppose $\delta$ returns $m' = \handover_i$.
    $\delta$ must also set $f_i'$ to a \emph{handover flag} indicating that $p$ has initiated a handover with its neighbor.
    \item On seeing $m' = \handover_i$ returned, the scheduler finds neighbor $q$ (the neighbor faced by the $i$-th port of $p$) and schedules $[q, p, q]$ as the next three particles to be activated.
    It activates $q$.
    \item Based on the inputs to $\delta$ for particle $q$, and in particular the handover flag $f_j$ from $p$ and the fact that it is contracted, $\delta$ must evaluate such that $q$ sets $f_j'$ as a \emph{will-expand flag}, sets $e' = j$, and sets $m' = \idle$.
    \item The scheduler now has $[p, q]$, so it activates $p$.
    \item Based on the inputs to $\delta$ for particle $p$, and in particular the will-expand flag $f_i$ from $q$, $\delta$ must evaluate such that it clears $f_i' = \eflag$ and sets $m' = \contract_i$ (setting $t' = \varepsilon$, following the rules above). Thus, it contracts.
    \item The scheduler now has $[q]$, so it activates $q$.
    \item Based on the inputs to $\delta$ for particle $q$, and in particular its expansion direction $e = j$, $\delta$ must evaluate such that $q$ clears $f_j' = \eflag$ and sets $m' = \expand_e$ (setting $t'$ to the corresponding tail port opposite $e$).
    \item The scheduler has no queued activations, so it chooses arbitrarily but fairly.
\end{enumerate}

A push handover initiated by a contracted particle $p$ with an expanded neighbor $q$ is handled similarly.
The first activation of $p$ is the same as Step 1 above, causing the scheduler to do the same queuing as in Step 2.
However, in Step 3, $q$ sees the handover flag but also that it is expanded, meaning this is a push handover.
Note, however, that a push handover is symmetric to a pull handover with the exception of which particle initiates; i.e., $p$ performing a push handover with $q$ yields the same result as $q$ performing a pull handover with $p$.
So, on seeing this is a push handover, $q$ simply proceeds as particle $p$ starting in Step 1, effectively exchanging roles with $p$.

The \emph{configuration} of a particle $p$ is $C(p) = (\vec{v}, o, q, t, e, (f_0, \ldots, f_9))$, where $\vec{v} \in V$ is the coordinates of its head node, $o \in O$ is its orientation, $q \in Q$ is its state, $t \in T$ is its tail port, $e \in E$ is its expansion direction, and each $f_i \in \Sigma$ is the flag on its $i$-th port, for $i \in \{0,1,\dots,9\}$.
Note that although the configuration of a particle $p$ includes all information needed to reconstruct $p$, particle $p$ itself does not have access to any global information or unique identifiers; in particular, it has no knowledge of $v$ or $o$.
The configuration of a particle system $P$ is $C^*(P) = \{C(p) : p \in P\}$, the set of all configurations of particles in $P$.
A system configuration is \emph{valid} if no two particles in the system occupy a common node in $\Gtri$.
We define $\mathcal{C}(P)$ to be the set of all valid system configurations of $P$.
An \emph{amoebot system} is defined as a 5-tuple $\mathcal{A} = (Q, \Sigma, \delta, P, \sigma)$, where $Q$ is a constant-size set of particle states, $\Sigma$ is a constant-size alphabet of flags, $\delta$ is the transition function, $P$ is the particle system, and $\sigma \in \mathcal{C}(P)$ is the initial system configuration of $\mathcal{A}$ mapping each particle to its starting configuration.


For system configurations $\alpha, \alpha' \in \mathcal{C}(P)$, where $\alpha \neq \alpha'$, we say $\alpha$ yields $\alpha'$ (denoted $\alpha \rightarrow^\mathcal{A} \alpha'$) if $\alpha$ can become $\alpha'$ after a single particle activation.
We use $\alpha \rightarrow^\mathcal{A}_* \alpha'$ if $\alpha$ yields $\alpha'$ in $0$ or more activations.
A sequence of configurations $(\alpha_0, \alpha_1, \ldots, \alpha_k)$ is a \emph{valid transition sequence} if for every $i \in [k]$ we have $\alpha_i \in \mathcal{C}(P)$, $\alpha_i \neq \alpha_{i+1}$, and $\alpha_i \to^\mathcal{A} \alpha_{i+1}$.
A configuration $\alpha \in \mathcal{C}(P)$ is called \emph{reachable} if there exists a valid transition sequence beginning at the initial configuration $\sigma$ and ending at $\alpha$.
A configuration $\alpha \in \mathcal{C}(P)$ is called \emph{terminal} if there is no configuration $\alpha' \in \mathcal{C}(P)$ such that $\alpha \to^\mathcal{A} \alpha'$.
A set of configurations $\Gamma \subseteq \mathcal{C}(P)$ is called \emph{terminal} if for all $\alpha \in \Gamma$ there is no configuration $\alpha' \not\in \Gamma$ such that $\alpha \to^\mathcal{A} \alpha'$ (i.e., no configuration in $\Gamma$ can transition to any configuration outside of $\Gamma$).
An amoebot system $\mathcal{A} = (Q, \Sigma, \delta, P, \sigma)$ is called \emph{directed} if every transition sequence from $\sigma$ leads to the same terminal configuration, or \emph{directed to set} $\Gamma$ if every transition sequence from $\sigma$ leads to a configuration in $\Gamma$.
Finally, given a shape $S$ (i.e., a connected set of nodes in $\Gtri$), we say that system $\mathcal{A}$ \emph{forms shape} $S$ if and only if, for some set of configurations $\Gamma \subseteq \mathcal{C}(P)$, $\mathcal{A}$ is directed to $\Gamma$ and for every $\alpha \in \Gamma$, the locations of the particles in $\alpha$ are exactly the locations of $S$ (up to translation and rotation).

\vspace{-10pt}
\section{Simulating Amoebot Systems with Tile Automata}\label{sec:simoverview}
\vspace{-5pt}

In this section, we present our main result, which is a construction that takes as input an amoebot system and which outputs a Tile Automata system that simulates it. However, we must first define what we mean by the term ``simulate'' in this context.

\vspace{-15pt}
\subsection{Defining simulation}
\vspace{-5pt}

\begin{wrapfigure}{r}{2.0in}
\vspace{-30pt}
    \centering
    \includegraphics[width=2.0in]{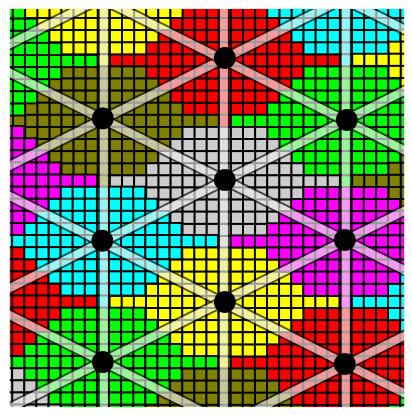}
    \caption{A portion of the tessellation by the macrotile shape of our construction (shown in Figures~\ref{fig:movement} and \ref{fig:tile_final}) with an overlay of $G_\Delta$.}
    \label{fig:macro-grid}
\vspace{-10pt}
\end{wrapfigure}

Intuitively, our simulation of an amoebot system by a Tile Automata system will consist of groups of tiles, called \emph{macrotiles}, which each represent a single amoebot particle. Starting from an assembly which maps (via a mapping function to be described) to the initial configuration of an amoebot system, singleton tiles as well as macrotiles will attach, detach, and change states. Any changes to the assembly, modulo a scale factor, will map to new, valid configurations of the amoebot system. Conversely, for any valid configuration change of the amoebot system, the assembly will be able to change in such a way that it represents the new amoebot configuration, under the mapping function. 

A \emph{macrotile} is a connected, finite region of the plane $\mathbb{Z}^2$, whose shape can be any polyomino composed of connected unit squares. For a macrotile shape $M$ to be valid to use for a simulation, it must tessellate. Since we are defining simulation of amoebot systems, which are embedded in the triangular grid, by Tile Automata, which are embedded in the square grid, a further condition is required for macrotile shapes. Let $\mathbb{T}(M)$ be a tessellation of the plane by macrotiles of shape $M$. Let $G$ be the graph formed where every node is a macrotile in $\mathbb{T}(M)$ and there is an edge between a pair of nodes if and only if they are adjacent to each other in $\mathbb{T}(M)$. Then, graph $G$ must be isomorphic to the triangular grid graph (i.e. the graph of the triangular grid where each intersection is a node). This means each macrotile has the same 6-neighbor neighborhood as nodes in the triangular grid graph (see Figure~\ref{fig:macro-grid}).

Let $\vec{v}$ be the coordinates of a node in $G_\Delta$, and let $m_{\vec{v}}$ be the macrotile location which corresponds to it. Given Tile Automata system $\Gamma$ and its set of producible assemblies $\texttt{PROD}_\Gamma$, for assembly $\A \in \texttt{PROD}_\Gamma$, let $\alpha \in A$ be a positioned assembly of $A$, and let $\alpha|m_{\vec{v}}$ be the (possibly empty) subassembly of $\alpha$ contained in the locations of $m_{\vec{v}}$. Given $\Gamma$ and an amoebot system $\mathcal{A} = (Q,\Sigma,\delta,P,\sigma)$, a \emph{macrotile representation function} $R$, from $\Gamma$ to $\mathcal{A}$, is a function which takes as input the portion of an assembly contained within a single macrotile locations, and which returns either information about the configuration of an amoebot particle from $P$, or $\epsilon$ (which maps to empty space). That is, given some $\alpha|m_{\vec{v}}$, $R(\alpha|m_{\vec{v}}) \in \{(t, e, o, q, (f_0, f_1, ..., f_9))$ $|$ $t \in T$, $o \in O$, $q \in Q$, $e \in E$, and $f_i \in \Sigma\} \cup \{\epsilon\}$, where $t \in T$ is the relative direction of a particle's tail from its head, $o \in O$ is its orientation offset, $q \in Q$ is its state, and each $f_i \in \Sigma$, for $0 \le i < 9$, is the flag in its $i$th port. An \emph{assembly representation function}, or simply \emph{representation function}, $R^*$ from $\Gamma$ to $\mathcal{A}$ takes as input an entire positioned assembly of $\Gamma$ and applies $R$ to every macrotile location and returns a corresponding amoebot system configuration from $\mathcal{C}(P)$.

For a positioned assembly $\alpha \in \texttt{PROD}_\Gamma$ such that $R^*(\alpha) = \alpha' \in \mathcal{C}(P)$, $\alpha$ is said to map \emph{cleanly} to $\alpha'$ under $R^*$ if for all non empty blocks $\alpha|m_{\vec{v}} \in \dom \alpha$, $\vec{v} \in \dom \alpha'$ or \sloppy $\vec{v}' \in \dom \alpha'$ for some $\vec{v}' = \vec{v} + \vec{u}$ where $\vec{u} \in \{(1,0),(0,1),(-1,0),(0,-1),(-1,1),(1,-1)\}$. In other words, $\alpha$ may have tiles in a macrotile location representing a particle in $\alpha'$, or empty space in $\alpha'$ but only if that position is adjacent to a particle in $\alpha'$.  We call such growth ``around the edges'' of $\alpha$ \emph{fuzz} and thus restrict it to be adjacent to macrotiles representing particles.

Note that the following definitions of \emph{follows}, \emph{models}, and \emph{simulates}, as well as the previous definitions of macrotiles, fuzz, etc. are based upon similar definitions used to prove results about simulation and intrinsic universality in \cite{IUSA,2HAMIU,IUNeedsCoop,WoodsMeunierSTOC} and several other papers.

\begin{definition}[$\mathcal{A}$ follows $\Gamma$]
Given Tile Automata system $\Gamma$, amoebot system $\mathcal{A}$, and assembly representation function $R^*$ from $\Gamma$ to $\mathcal{A}$, we say that $\mathcal{A}$ \emph{follows} $\Gamma$ (under $R$), and we write $\mathcal{A} \dashv_R \Gamma$, if $\alpha \rightarrow^\Gamma \beta$, for $\alpha,\beta \in \texttt{PROD}_\Gamma$, implies that $R^*(\alpha) \rightarrow^\mathcal{A}_* R^*(\beta)$.
\end{definition}

\begin{definition}[$\Gamma$ models $\mathcal{A}$]

Given Tile Automata system $\Gamma$, amoebot system $\mathcal{A} = (Q,\Sigma,\delta,P,\sigma)$, and assembly representation function $R^*$ from $\Gamma$ to $\mathcal{A}$, we say that $\Gamma$ \emph{models} $\mathcal{A}$ (under $R$), and we write $\Gamma \models_R \mathcal{A}$, if for every $\alpha \in \mathcal{C}(P)$, there exists $\Psi \subset \texttt{PROD}_\Gamma$ where $R^*(\alpha') = \alpha$ for all $\alpha' \in \Psi$, such that, for every $\beta \in \mathcal{C}(P)$ where $\alpha \rightarrow^\mathcal{A} \beta$, (1) for every $\alpha' \in \Psi$ there exists $\beta' \in \texttt{PROD}_\Gamma$ where $R^*(\beta') = \beta$ and $\alpha' \rightarrow^\Gamma \beta'$, and (2) for every $\alpha'' \in \texttt{PROD}_\Gamma$ where $\alpha'' \rightarrow^\Gamma \beta'$, $\beta' \in \texttt{PROD}_\Gamma$, $R^*(\alpha'') = \alpha$, and $R^*(\beta') = \beta$, there exists $\alpha' \in \Psi$ such that $\alpha' \rightarrow^\Gamma \alpha''$.
\end{definition}

Definition 2 essentially specifies that every time $\Gamma$ simulates an amoebot configuration $\alpha \in \mathcal{C}(P)$, there must be at least one valid growth path in $\Gamma$ for each of the possible next configurations that $\alpha$ could transition into from $\alpha$, which results in an assembly in $\Gamma$ that maps to that next step.

\begin{definition}[$\Gamma$ simulates $\mathcal{A}$]
Given Tile Automata system $\Gamma$, amoebot system $\mathcal{A}$, and assembly representation function $R^*$ from $\Gamma$ to $\mathcal{A}$, if $\mathcal{A} \dashv_R \Gamma$ and $\Gamma \models_R \mathcal{A}$, we say that $\Gamma$ \emph{simulates} $\mathcal{A}$ under $R$.
\end{definition}

With the definition of what it means for a Tile Automata system to simulate an amoebot system, we can now state our main result.

\begin{theorem}\label{thm:TAsimAS}
Let $\mathcal{A}$ be an arbitrary amoebot system. There exists a Tile Automata system $\Gamma$ and assembly representation function $R^*$ from $\Gamma$ to $\mathcal{A}$ such that $\Gamma$ simulates $\mathcal{A}$ under $R$. Furthermore, the simulation is at scale factor 100.
\end{theorem}

To prove Theorem~\ref{thm:TAsimAS}, we let $\mathcal{A} = (Q,\Sigma_\mathcal{A},\delta,P,\sigma)$ be an arbitrary amoebot system. We will now show how to construct a Tile Automata system $\Gamma = (\Sigma_\Gamma,\Lambda,\Pi,\Delta,\tau)$ such that $\Gamma$ simulates $\mathcal{A}$ at scale factor 100. The rest of this section contains details of our construction.

\vspace{-5pt}
\subsection{Construction definitions} \label{constuctionDefs}


\emph{\textbf{Neighborhood}} - In the geometric amoebots model, particles are aware of the occupation of all locations on the lattice adjacent to their own. The neighborhood of a given location on the lattice is the set of its six neighbors. Pertaining to a particle, we say that a particle's neighborhood is the set all particles occupying adjacent locations on the lattice, defined by $N(p)$, where \emph{p} is a particle. Note that $|N(p)| \leq 6$ if $p$ is contracted and $10$ if it is expanded.

\emph{\textbf{Macrotile}} - A $\tau$-stable assembly of TA tiles $g$ such that the macrotile representation function $R$ maps $g$ to a valid particle in $\calA$. This simulation makes use of macrotiles with an approximately hexagonal shape and special tiles within each macrotile used to calculate information about its movement and neighborhood. See Figure~\ref{fig:tile_final} for an overview.

\emph{\textbf{Clock Tiles}} - The tiles at the middle of every particle macrotile used to keep track of state, flags, $T$ value, and neighborhood information. The middle clock tile is responsible for maintaining the particle's state $q \in Q$, and the surrounding clock tiles (called subordinate clock tiles) combine information from the particle edges and neighbors to pass into the central clock tile. 

\emph{\textbf{Wire Tiles}} - Rows of tiles leading from the bank of clock tiles in the middle of every macrotile to each edge, purposed with transmitting information from the clock to the neighboring tiles and available edges.

\emph{\textbf{Filler Tiles}} - Tiles that serve no function within a macrotile other than to maintain connectivity with other components and shape the macrotile.

\emph{\textbf{Flag Tiles}} - Exposed wire ends on each side of the tile responsible for maintaining flag states from $\Sigma$ in $\mathcal{A}$, as well as reading flags from their respective neighbors. Neighboring flags are retrieved via wire transmission.

\emph{\textbf{Timing Tiles}} - Individual tiles in $\Gamma$ that diffuse into specific slots in the particle macrotiles that ``start" that particle's turn, and disconnect after the turn is finished. Timing tiles are inert except for connecting to a central clock tile, and serve as the ``asynchronous clock" for our simulation. 

\emph{\textbf{$\epsilon$ Tiles}} - Individual tiles in $\Gamma$ that attach to the available flag tiles of macrotiles with non-full neighborhoods who are querying or attempting to lock their neighborhood flags. $\epsilon$ tiles can only attach to the exposed end of a wire displaying a lock or query signal flag. After attaching, they serve only to undergo a single state transition, which indicates to the wire end that there exists no neighbor there. After this transition, the wire propagates this information back to its clock bank and the $\epsilon$ tile detaches.

\emph{\textbf{Floating Macrotile}} - These macrotiles (``floats"), will represent (portions of) particles in $\calA$ but are not connected to the swarm. They attach to valid sites along the perimeter and simulate the ``head" of an expanding particle. 

\emph{\textbf{Configuration Tile}} - After macrotiles complete their turn, they combine all values of their configuration of which they are aware $(q \in Q, e \in E, (f_0, f_1, ..., f_9 \in \Sigma^{10}), t \in T)$ into a single tile proximal to the clock bank called the \textit{Configuration Tile}. When a macrotile is engaging in an expansion or handover move, the configuration tile is used by the representation function $R$ to map the active macrotile to its previous configuration, until the transition completes. An active macrotile engaged in these moves will be mapped by its configuration tile until it's no longer displaying a flag indicating it is engaged in either an expansion or handover on any of its wires, ensuring the simulation of an atomic move. For a more complete explanation, see appendix B, simulation details. 

\subsection{Simulation Overview}

\begin{wrapfigure}{r}{2.0in}
\centering
\vspace{-30pt}
\includegraphics[width=2.0in]{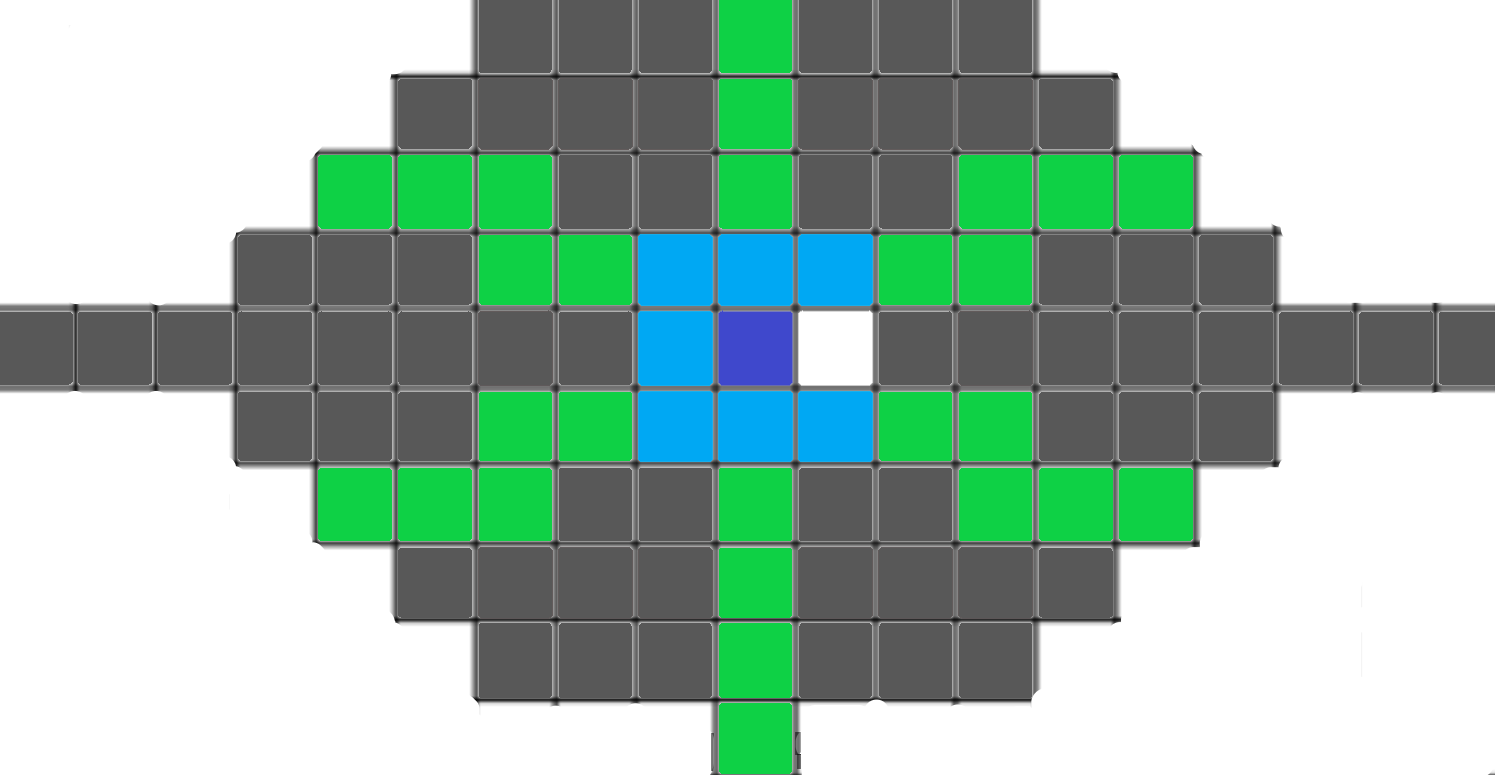}
\caption{Blue tiles are subordinate clock tiles, dark blue is the central clock tile, green tiles wires, and grey tiles filler. The empty location immediately east of the central clock tile is where the timing tile attaches to signal the central clock tile to begin the turn. Flags are displayed on the the outermost tile of every wire.}
\vspace{-20pt}
\label{fig:tile_final}
\end{wrapfigure}

The simulation of $\calA$ by $\Gamma$ is affected with the assistance of the hexagonal macrotiles and the signals implemented via TA state transitions. In \cite{freezing}, since there were only four directions from which signals could come, it was sufficient for each macrotile to have one clock tile, which would transition its own state based on signals received from wires and send its state down the wires. Since the geometric amoebots model exists on the $G_\Delta$, signals can come from up to six directions, necessitating the use of multiple clock tiles. Figure~\ref{fig:tile_final} contains a high-level depiction of a macrotile in $\Gamma$ that simulates a particle of $\calA$. 

\begin{figure}[h]
\centering
\includegraphics[width=3.5in]{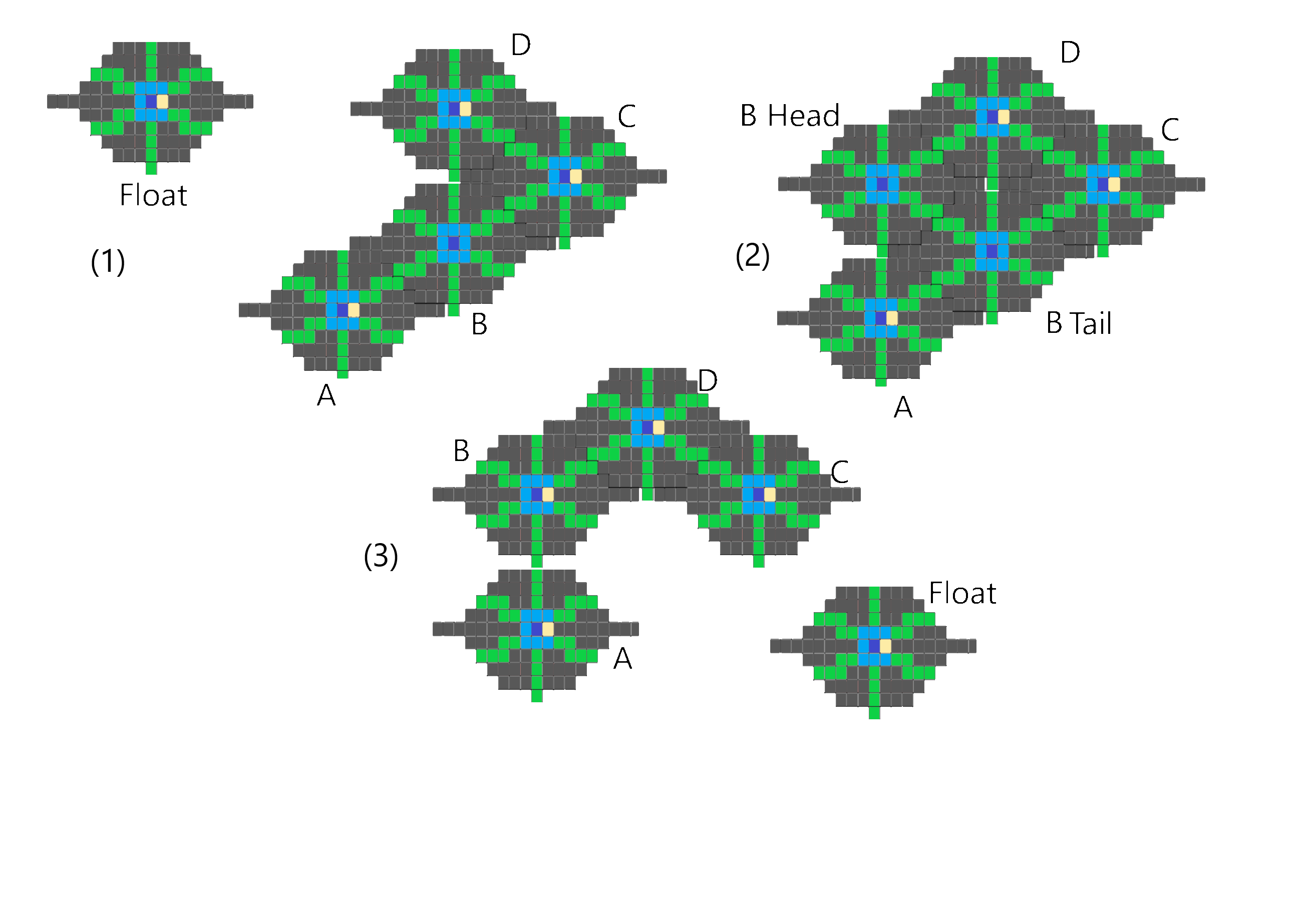}
\vspace{-25pt}
\caption{Particle B simulates movement by allowing a float to attach to a free edge, then detaching the original macrotile representing particle B over the course of two activations.}
\label{fig:movement}
\end{figure}

Figure~\ref{fig:movement} illustrates a simple example of simulated particle  movement. Macrotiles must be initially arranged into a configuration $\alpha$ that under $R^*(\alpha)$ maps to a valid configuration $\alpha' \in \mathcal{C}(P)$, with connected edges representing adjacency in $\alpha'$. Macrotiles start with their respective states, flags, and $t$ values set to whatever those states are for the corresponding particle in $\alpha'$. Swarm macrotiles may then begin to accept timing tiles, starting their turns. We use neighborhood \emph{lock signals} to ensure that no particles that are in the same neighborhood attempt to move at the same time, avoiding asynchronous conflicts. Expansion is facilitated by the attachment along perimeter sites of floating macrotiles. The authors additionally considered systems where moves progressed by growing a new macrotile wherever a particle wanted to expand, but this construction technique requires a longer wait between neighborhood locks and unlocks. In the interest of minimizing the overhead that simulation requires, we wanted to minimize the amount of time that a neighborhood had to be locked in order to encourage collaborative movement, and decided to use prefabricated floating macrotiles.

Once a particle macrotile has received a timing tile, it only continues its turn if it is not already locked by a neighbor. If the particle is not locked down, then the active macrotile sends signals to all of its neighbors to lock down its neighborhood and it can continue without fear of causing conflict. Should two $lock$ signals be traveling towards each other along a shared wire between two macrotiles, whichever signal is first carried into the other macrotile's wires via state transitions overwrites the signal originating from the slower macrotile, and the faster propagating signal's originator locks down the slower. The neighboring flags are needed to simulate the transition function, and are sent from neighbors to the active macrotile via wire transmission. Once the active macrotile has decided its new state, flags, and move, it updates this information and attempts to execute its chosen move. If the move is a simple expansion, it marks the site where it wants to expand with a valid attachment signal and keeps the neighborhood locked until a floating macrotile connects to it, representing the ``head" of the expanding particle. If the particle chooses to contract, it sends signals to the tail to detach from the swarm, whereupon it will become another float, and then unlocks its neighborhood. If the particle chose a $\handover_i$, it performs some additional checks to ensure viability and then sends signals either ``taking over" one of a neighboring expanded particle's macrotiles, or ceding control of one of its macrotiles to a neighboring tile. In the case of $\handover_i$, an additional state and flags for the subordinate particle are returned from the transition function, to be propagated from the active macrotile to the subordinate macrotile via wire transmissions as control of the macrotile changes hands. 



\vspace{-5pt}



\vspace{-5pt}
\section{Simulation of Movement} \label{sec:simmove}
\vspace{-5pt}

Before the simulation begins, given $\mathcal{A}$ with initial configuration $\sigma$, we produce $\Lambda$ for $\Gamma$  which consists of a connected configuration of macrotiles, each macrotile mapping to a corresponding particle in $\sigma$ under $R$. To capture the asynchronous nature of amoebot particle activations, we utilize timing tile diffusion into macrotiles to ``activate" them for their turns. After attachment, the tile sends a signal to the central clock to start its turn. After a given macrotile has started its turn and has successfully locked its neighborhood, it gathers all information necessary for its transition function via wire-propagated signals (detailed in the technical appendix). To ensure that the transition function affected by macrotiles is isomorphic to the transition function affected by $\mathcal{A}$, we combine all of the flags, $e$ and $t$ values (resulting in a $|\Sigma^{10}| * |t| * |e|$ increase in state complexity for clock tiles) into the tile to the left of the central clock tile. Once this value is at that tile, the central clock tile, which holds the state, undergoes a state transition defined in the construction of $\Delta$.

Once the new state, flags, and move are produced, the particle propagates its flags down their respective wires and in the case that $\handover_i$ is returned for the value of $m$, the particle additionally sends a signal (detailed in the technical appendix) to its $i$th neighbor to ensure that the neighbor has the proper orientation to facilitate that move. For $\expand_i$, the active macrotile sends a signal down the wire in the $i$th direction that allows a float to attach to that edge. After connection of a float, the active macrotile further sends a $Copy Signal$ to the newly attached float so the new float can copy the states and relevant flags of the expanding macrotile and fully become the head. The float sends an $Acknowledgement Signal$ after it is displaying the proper state and flags, which tells the newly expanded macrotile that it's safe to unlock its neighborhood. For $\contract_i$, the expanded macrotile sends a $Detach$ signal to the macrotile that contains the $i$th port. After detachment, the recently contracted macrotile unlocks its neighborhood. For $\idle$, states and flags may be updated, but no change to the orientation of the simulated particles occurs. Once a macrotile executing an idle move changes its state and sends the new flags to its flag tiles, it unlocks its neighborhood and ends its turn. 


\vspace{-5pt}
\subsection{\handover} \label{Handovers}

\begin{figure}[h]
\centering
\vspace{-5pt}
\includegraphics[width=2.0in]{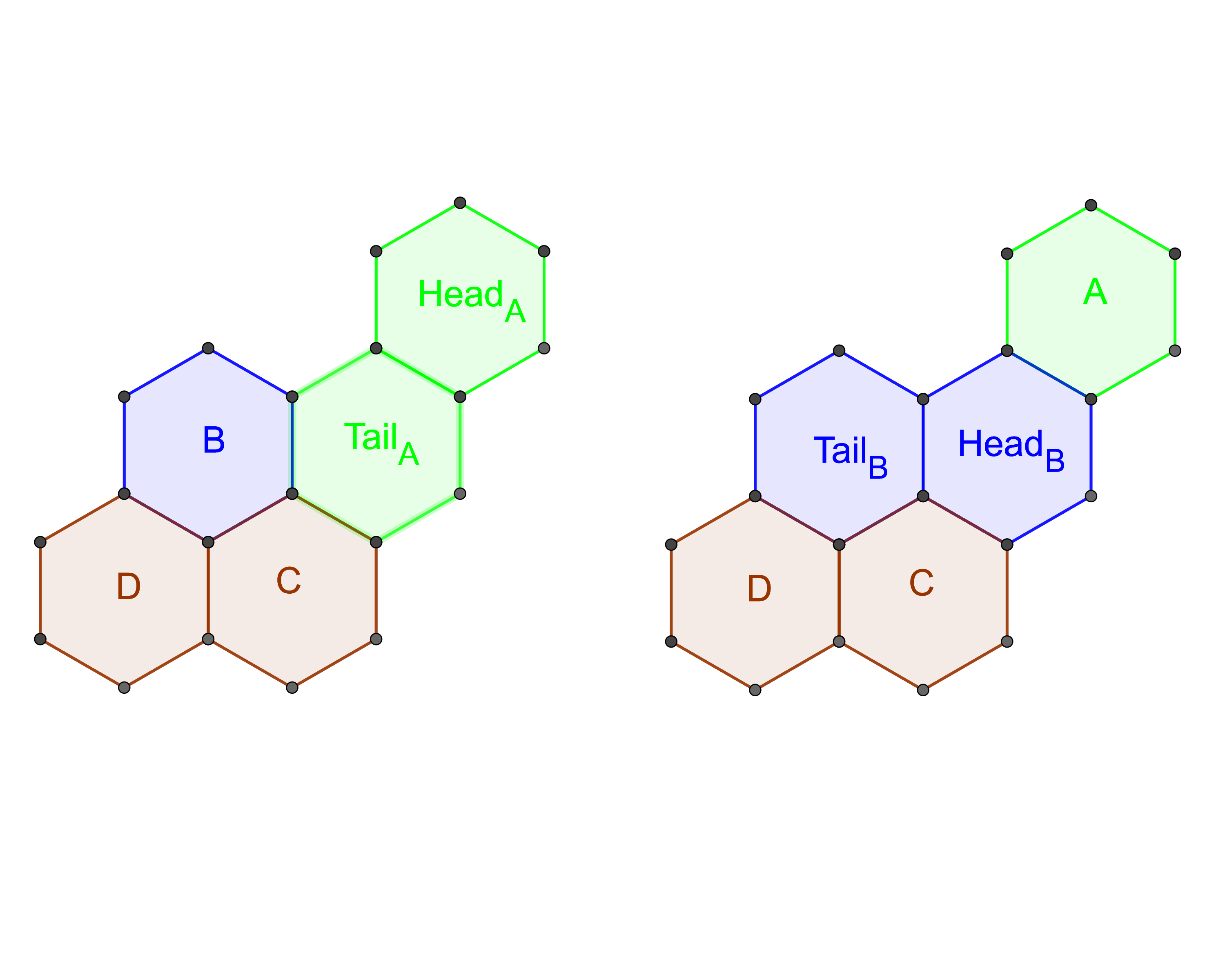}
\vspace{-20pt}
\caption{A $handoverPull_i$ movement executed by particle A. The overall orientation of the swarm does not change, but $Tail_A$ changes hands and becomes the head of B, the subordinate particle in this exchange.}
\vspace{-15pt}
\label{fig:handover}
\end{figure}

The Amoebots model defines a move called the $\handover_i$, which allows two neighboring particles, one expanded and one contracted, to simultaneously contract from and expand into the same grid location. This move can be initiated by either particle, and involves the initiating particle's neighbor in the $ith$ direction. There are four possibilities for handover contracts: A contracted particle can expand into a spot previously occupied by a head, a contracted particle can expand into a spot previously occupied by a tail, an expanded particle can contract and ``force" its neighbor into expanding into wherever its head is currently located, and an expanded particle can contract and ``force" its neighbor into expanding into wherever its tail is currently located. It's necessary that both moves happen simultaneously to enable certain actions such as moving a collection of particles through a static tunnel of width one. For any $handoverPull$ movement, the initiating particle must be expanded and the subordinate particle contracted, while for any $handoverPush$ movement, the reverse is true.

All moves involve locking down the active particle's neighborhood. Since $\handover_i$ necessarily changes the orientation of both the initiating and subordinate particles, it is necessary to lock down the respective neighborhoods of all particles involved. Thus, the $handoverPull_i$ movement requires a $Progressive Lockdown Signal_i$ to be sent to the subordinate particle(s) to be further propagated to their respective neighborhoods. Note, that for the duration of the execution of a handover move, a $\handover_i$ flag will be displayed, ensuring that the macrotile maps to its previous configuration under the representation function $R$ until the $\handover_i$ or $\expand_i$ flags are cleared from the macrotiles' respective ports. This ensures that we have smooth transitions for macrotiles mapping to the atomic transitions of amoebots.

Handover moves in our simulation occur after the initiating particle has locked down its own neighborhood, checked to ensure move validity with the $\handover_i$ signal, and further sent a $Progressive Lockdown Signal_i$ to each subordinate particle. The initiating particle ensures that the configuration of expansions and contractions are appropriate for the move it is attempting, and then sends a signal to the macrotile that will be passed from one particle to another. This signal contains the new state and flags for the subordinate  macrotile(s). Once these fields are updated, and the $\handover_i$ flags are cleared from the macrotiles' wires, the macrotile that changed state is considered to be the head of the newly expanded particle, which sends an acknowledgement signal back to the initiating macrotile. Macrotiles engaged in a $\handover_i$ move are mapped by $R$ via their configuration tiles until they clear the flags, after which they revert to the normal mode of operation for $R$, which checks a macrotile's central clock tile and the subordinate clock tile to the immediate west. After receiving the acknowledgement of successful transition, the initiating macrotile sends out unlock signals to its neighborhood (which necessarily propagate from subordinate particles to their respective neighborhoods as well), and finally the initiating macrotile detaches its timing tile, ending the turn.

\vspace{-10pt}
\subsection{Attachment Sites}
\vspace{-5pt}

To avoid potential conflicts between tiles undergoing their state transitions and the floating macrotiles, we only allow the floating macrotiles to attach to valid attachment sites along the perimeter of the swarm. The perimeter of the swarm shares no affinity with the float tiles by default. Only after a perimeter particle undergoes its transition function and returns $M = \expand_i$ and marks that edge with a state that has affinity with the float can any float attach to the particle that wants to expand. Floats attach to the swarm via a $\tau$-strength attachment along the exposed wire end of the perimeter particle that is attempting to expand. The only time macrotiles can detach from the swarm is when a given expanded macrotile receives a $Detach$ signal from its other end. When this occurs, the detaching macrotile sends signals from its clock bank to all edges to have them change their exposed wire ends to a state with no affinity to the swarm. Once the macrotile is no longer part of the swarm, it sends $query$ signals to all of its edges. In the case that a macrotile receives six $\epsilon$ responses from its wires (that is, it has no neighbors) after a query signal, it undergoes states transitions in the clock and wires that make it a float. The newly contracted macrotile prevents unintentional reattachments of other floats because the $Detach$ signal, after leaving the contracted particle's wire, leaves the wire in its respective flag state with no affinity with floats. The wire that corresponds to the detached macrotile will not allow attachments again until it receives an $Attach$ signal from the central clock bank again. 


\vspace{-10pt}
\section{Conclusion} \label{sec:conclude}
\vspace{-10pt}

We have presented a simulation construction in which an amoebots system can be simulated by a collection of Tile Automata macrotiles. The mechanisms by which particle movement is simulated were discussed as well, such as how the atomic actions of the amoebots model were replicated within the simulation without threat of interruption via state transitions. We hope this fits into a larger schema of comparing the power of various computational models by simulation.

\section*{Acknowledgements}

The authors would like to thank Schloss Dagstuhl – Leibniz Center for Informatics and the organizers and participants of Dagstuhl Seminar 18331 ``Algorithmic Foundations of Programmable Matter'' \cite{Dagstuhl2018}. The initial brainstorming and work for this paper began during that workshop and was inspired by many interesting discussions with the participants.

\vspace{-5pt}

\bibliographystyle{splncs04}
\bibliography{selfAssembly,amoebots,tam}

\begin{thebibliography}{10}
\providecommand{\url}[1]{\texttt{#1}}
\providecommand{\urlprefix}{URL }
\providecommand{\doi}[1]{https://doi.org/#1}

\bibitem{Arroyo2018}
Andr{\'e}s~Arroyo, M., Cannon, S., Daymude, J.J., Randall, D., Richa, A.W.: A
  stochastic approach to shortcut bridging in programmable matter. Natural
  Computing  \textbf{17}(4),  723--741 (2018)

\bibitem{Dagstuhl2018}
Berman, S., Fekete, S.P., Patitz, M.J., Scheideler, C.: { Algorithmic
  Foundations of Programmable Matter (Dagstuhl Seminar 18331)}. Dagstuhl
  Reports  \textbf{8}(8),  48--66 (2019). \doi{10.4230/DagRep.8.8.48},
  \url{http://drops.dagstuhl.de/opus/volltexte/2019/10235}

\bibitem{Cannon2016}
Cannon, S., Daymude, J.J., Randall, D., Richa, A.W.: A {M}arkov chain algorithm
  for compression in self-organizing particle systems. In: Proceedings of the
  2016 ACM Symposium on Principles of Distributed Computing. pp. 279--288. PODC
  '16, ACM, New York, NY, USA (2016)

\bibitem{freezing}
Chalk, C., Luchsinger, A., Martinez, E., Schweller, R., Winslow, A., Wylie, T.:
  Freezing simulates non-freezing tile automata. In: DNA Computing and
  Molecular Programming. pp. 155--172. Springer International Publishing (2018)

\bibitem{Daymude2017}
Daymude, J.J., Gmyr, R., Richa, A.W., Scheideler, C., Strothmann, T.: Improved
  leader election for self-organizing programmable matter. In: Algorithms for
  Sensor Systems. pp. 127--140. ALGOSENSORS '17, Springer, Cham (2017)

\bibitem{Daymude2019}
Daymude, J.J., Hinnenthal, K., Richa, A.W., Scheideler, C.: Computing by
  programmable particles. In: Distributed Computing by Mobile Entities: Current
  Research in Moving and Computing, pp. 615--681. Springer, Cham (2019)

\bibitem{2HAMIU}
Demaine, E.D., Patitz, M.J., Rogers, T.A., Schweller, R.T., Summers, S.M.,
  Woods, D.: The two-handed assembly model is not intrinsically universal. In:
  40th International Colloquium on Automata, Languages and Programming, ICALP
  2013, Riga, Latvia, July 8-12, 2013. Lecture Notes in Computer Science,
  Springer (2013)

\bibitem{Derakhshandeh2016}
Derakhshandeh, Z., Gmyr, R., Richa, A.W., Scheideler, C., Strothmann, T.:
  Universal shape formation for programmable matter. In: Proceedings of the
  28th ACM Symposium on Parallelism in Algorithms and Architectures. pp.
  289--299. SPAA '16, ACM, New York, NY, USA (2016)

\bibitem{Derakhshandeh2017}
Derakhshandeh, Z., Gmyr, R., Richa, A.W., Scheideler, C., Strothmann, T.:
  Universal coating for programmable matter. Theoretical Computer Science
  \textbf{671},  56--68 (2017)

\bibitem{amoebots}
Derakhshandeh, Z., Richa, A., Dolev, S., Scheideler, C., Gmyr, R., Strothmann,
  T.: Brief announcement: Amoebot-a new model for programmable matter. In:
  Annual ACM Symposium on Parallelism in Algorithms and Architectures. pp.
  220--222. Association for Computing Machinery (2014)

\bibitem{IUSA}
Doty, D., Lutz, J.H., Patitz, M.J., Schweller, R.T., Summers, S.M., Woods, D.:
  The tile assembly model is intrinsically universal. In: Proceedings of the
  53rd Annual IEEE Symposium on Foundations of Computer Science. pp. 302--310.
  FOCS 2012 (2012)

\bibitem{Polyominoes}
Fekete, S.P., Hendricks, J., Patitz, M.J., Rogers, T.A., Schweller, R.T.:
  Universal computation with arbitrary polyomino tiles in non-cooperative
  self-assembly. In: Proceedings of the Twenty-Sixth Annual ACM-SIAM Symposium
  on Discrete Algorithms (SODA 2015), San Diego, CA, USA {\rm January 4-6,
  2015}. pp. 148--167 (2015)

\bibitem{Polygons}
Gilbert, O., Hendricks, J., Patitz, M.J., Rogers, T.A.: Computing in continuous
  space with self-assembling polygonal tiles. In: Proceedings of the
  Twenty-Seventh Annual ACM-SIAM Symposium on Discrete Algorithms (SODA 2016),
  Arlington, VA, USA {\rm January 10-12, 2016}. pp. 937--956 (2016)

\bibitem{Signals3D}
Hendricks, J., Padilla, J.E., Patitz, M.J., Rogers, T.A.: Signal transmission
  across tile assemblies: 3{D} static tiles simulate active self-assembly by
  2{D} signal-passing tiles. In: Soloveichik, D., Yurke, B. (eds.) DNA
  Computing and Molecular Programming. Lecture Notes in Computer Science,
  vol.~8141, pp. 90--104. Springer International Publishing (2013)

\bibitem{jDuples}
Hendricks, J., Patitz, M.J., Rogers, T.A., Summers, S.M.: The power of duples
  (in self-assembly): It's not so hip to be square. Theoretical Computer
  Science  (2015)

\bibitem{KaoSchweller08}
Kao, M.Y., Schweller, R.T.: Randomized self-assembly for approximate shapes.
  In: Aceto, L., Damg{\aa}rd, I., Goldberg, L.A., Halld{\'o}rsson, M.M.,
  Ing{\'o}lfsd{\'o}ttir, A., Walukiewicz, I. (eds.) ICALP (1). Lecture Notes in
  Computer Science, vol.~5125, pp. 370--384. Springer (2008)

\bibitem{IUNeedsCoop}
Meunier, P.E., Patitz, M.J., Summers, S.M., Theyssier, G., Winslow, A., Woods,
  D.: Intrinsic universality in tile self-assembly requires cooperation. In:
  Proceedings of the ACM-SIAM Symposium on Discrete Algorithms (SODA 2014),
  (Portland, OR, USA, January 5-7, 2014). pp. 752--771 (2014)

\bibitem{WoodsMeunierSTOC}
Meunier, P., Woods, D.: The non-cooperative tile assembly model is not
  intrinsically universal or capable of bounded turing machine simulation. In:
  Proceedings of the 49th Annual {ACM} {SIGACT} Symposium on Theory of
  Computing, {STOC} 2017, Montreal, QC, Canada, June 19-23, 2017. pp. 328--341
  (2017)

\bibitem{jSignals}
Padilla, J.E., Patitz, M.J., Schweller, R.T., Seeman, N.C., Summers, S.M.,
  Zhong, X.: Asynchronous signal passing for tile self-assembly: Fuel efficient
  computation and efficient assembly of shapes. International Journal of
  Foundations of Computer Science  \textbf{25}(4),  459--488 (2014)

\bibitem{SingleNegative}
Patitz, M.J., Schweller, R.T., Summers, S.M.: Exact shapes and turing
  universality at temperature 1 with a single negative glue. In: Proceedings of
  the 17th international conference on DNA computing and molecular programming.
  pp. 175--189. DNA'11 (2011)

\bibitem{Winf98}
Winfree, E.: Algorithmic Self-Assembly of {D}{N}{A}. Ph.D. thesis, California
  Institute of Technology (June 1998)

\bibitem{WoodsIUSurvey}
Woods, D.: Intrinsic universality and the computational power of self-assembly.
  Philosophical Transactions of the Royal Society of London A: Mathematical,
  Physical and Engineering Sciences  \textbf{373}(2046) (2015)

\bibitem{Nubots}
Woods, D., Chen, H.L., Goodfriend, S., Dabby, N., Winfree, E., Yin, P.: Active
  self-assembly of algorithmic shapes and patterns in polylogarithmic time. In:
  Proceedings of the 4th conference on Innovations in Theoretical Computer
  Science. pp. 353--354. ITCS '13, ACM, New York, NY, USA (2013)

\end{thebibliography}

\pagebreak

\appendix

\section*{Technical Appendix}

\section{The Tile Automata Model}
\subsection{States, tiles, and assemblies}

\textbf{Tiles and States}
Let $\Sigma$ be an alphabet of \emph{state types}. A tile $t$ is a unit square centered at a point of the discrete plane, denoted by $L(t) \in \mathbb{Z}^2$. Each tile is assigned a state, $S(t) \in \Sigma$. Two tiles $t_1$ and $t_2$ are said to have the same \emph{tile type} if $S(t_1) = S(t_2)$.

\textbf{Affinity Function} Let $D = \{\perp, \vdash\}$, where $\perp$ and $\vdash$ represent above-below and side-by-side orientations of a pair of tiles, respectively. An \emph{affinity function} $\Pi : \Sigma^2 \times D \rightarrow \mathbb{N}$, where the output is the \emph{affinity strength} between tiles with the input pair of states and the relative positions specified by the given direction $d \in D$.

\textbf{Transition Rules} Transition rules allow the states of tiles to change based on their neighbors. A \emph{transition rule} is a 5-tuple $(S_{ia},S_{2a},S_{1b},S_{2b},d)$, with each $S_{ia},S_{2a},S_{1b},S_{2b} \in \Sigma$ and $d \in D$, that specifies that if two tiles are in states $S_{1a}$ and $S_{2a}$ and adjacent to each other in orientation $d$, then they can transition into states $S_{2a}$ and $S_{2b}$, respectively. A transition rule is considered to be a \emph{single-transition rule} if either $S_{1a} = S_{1b}$ or $S_{2a} = S_{2b}$, and a \emph{double-transition rule} otherwise.



\textbf{Assemblies} A \emph{positioned shape} is any connected subset of $\mathbb{Z}^2$. A \emph{positioned assembly} is a set of tiles at unique coordinates in $\mathbb{Z}^2$, which can be represented as a mapping from locations to tile states (or empty locations), $\alpha: \mathbb{Z}^2 \rightarrow \Sigma \cup \{\epsilon\}$. The positioned shape of a positioned assembly $\alpha$ is the set of coordinates of its tiles, denoted as $\texttt{SHAPE}_\alpha$. For a positioned assembly $\alpha$, let $\alpha(x,y)$ denote the state of the tile with location $(x,y) \in \mathbb{Z}^2$ in $\alpha$.

For a given positioned assembly $\alpha$ and affinity function $\Pi$, define the \emph{bond graph} $G_\alpha$ to be the weighted grid graph in which:
\begin{enumerate}
    \item each tile of $\alpha$ is a vertex
    \item no edge exists between non-adjacent tiles
    \item the weight of an edge between adjacent tiles $t_1$ and $t_2$ with locations $(x_1,y_1)$ and $(x_2,y_2)$, respectively, is:
    \begin{enumerate}
        \item $\Pi(S(t_1),S(t_2),\perp)$ if $y_1 > y_2$
        \item $\Pi(S(t_2),S(t_1),\perp)$ if $y_1 < y_2$
        \item $\Pi(S(t_1),S(t_2),\vdash)$ if $x_1 < x_2$
        \item $\Pi(S(t_2),S(t_1),\vdash)$ if $x_1 > x_2$
    \end{enumerate}
\end{enumerate}

A positioned assembly $\alpha$ is said to be $\tau$-stable for positive integer $\tau$ provided the bond graph $G_{\alpha}$ has min-cut at least $\tau$.

For a positioned assembly $\alpha$ and integer vector $\vec{v} = (v_1,v_2)$, let $\alpha_{\vec{v}}$ denote the positioned assembly obtained by translating each tile in $\alpha$ by $\vec{v}$. An \emph{assembly} is a set of all translations $\alpha_{\vec{v}}$ of a positioned assembly $\alpha$. A \emph{shape} is the set of all integer translations for some subset of $\mathcal{Z}^2$, and the shape of an assembly $A$ is defined to be the set of positioned shapes of all positioned assemblies in $A$. The \emph{size} of either an assembly or shape $X$, denoted as $|X|$, refers to the number of tiles in any positioned assembly of $X$.

\textbf{Breakable Assemblies} An assembly is $\tau$-\emph{breakable} if it can be split into two assemblies along a cut whose total affinity strength sums to less than $\tau$. Formally, an assembly $C$ is \emph{breakable} into assemblies $A$ and $B$ if the bond graph $G_C$ for some positioned assembly $\gamma \in C$ has a cut $(\alpha,\beta)$ for positioned assemblies $\alpha \in A$ and $\beta \in B$ of affinity strength less than $\tau$. We call assemblies $A$ and $B$ \emph{pieces} of the breakable assembly $C$.

\textbf{Combinable Assemblies} Two assemblies are $\tau$-\emph{combinable} provided they may attach along a border whose strength sums to at least $\tau$. Formally, two assemblies $A$ and $B$ are $\tau$-combinable into an assembly $C$ provided that $G_\gamma$ for any $\gamma \in C$ has a cut $(\alpha,\beta)$ of strength at least $\tau$ for some positioned assemblies $\alpha \in A$ and $\beta \in B$. We call $C$ a \emph{combination} of $A$ and $B$.

\textbf{Transitionable Assemblies} Let $\Delta$ be a set of transition rules. An assembly $A$ is \emph{transitionable}, with respect to $\Delta$, into assembly $B$ if and only if there exist $\alpha \in A$ and $\beta \in B$ such that for some pair of adjacent tiles $t_i,t_j \in \alpha$:
\begin{enumerate}
    \item $\exists$ a pair of adjacent tiles $t_h,t_k \in \beta$ with $L(t_i) = L(t_h)$ and $L(t_j) = L(t_k)$
    \item $\exists$ a transition rule $\delta$ in $\Delta$ such that $\delta = (S(t_i),S(t_j),S(t_h),S(t_k),\perp)$ or $\delta = (S(t_i),S(t_j),S(t_h),S(t_k),\vdash)$
    \item $\mathcal{A} - \{t_i,t_j\} = \mathcal{B} - \{t_h,t_k\}$
\end{enumerate}

\subsection{Tile Automata (TA) model}

A \emph{tile automata system} is a 5-tuple $(\Sigma,\Pi,\Lambda,\Delta,\tau)$ where $\Sigma$ is an alphabet of state types, $\Pi$ is an affinity function, $\Lambda$ is a set of initial assemblies with each tile assigned a state from $\Sigma$, $\Delta$ is a set of transition rules for states in $\Sigma$, and $\tau \in \mathbb{N}$ is the \emph{stability threshold}. When the affinity function and state types are implied, we let $(\Lambda,\Delta,\tau)$ denote a tile automata system. An example tile automata system can be seen in Figure~\ref{fig:TA_ex}.

\begin{definition}[Tile Automata Producibility]
For a given tile automata system $\Gamma = (\Sigma,\Lambda,\Pi,\Delta,\tau)$, the set of \emph{producible assemblies} of $\Gamma$, denoted $\texttt{PROD}_\Gamma$, is defined recursively:
\begin{enumerate}
    \item (Base) $\Lambda \subseteq \texttt{PROD}_\Gamma$
    \item (Recursion) Any of the following:
    \begin{enumerate}
        \item (Combinations) For any $A,B \in \texttt{PROD}_\Gamma$ such that $A$ and $B$ are $\tau$-combinable into $C$, then $C \in \texttt{PROD}_\Gamma$
        \item (Breaks) For any $C \in \texttt{PROD}_\Gamma$ such that $C$ is $\tau$-breakable into $A$ and $B$, then $A,B \in \texttt{PROD}_\Gamma$
        \item (Transitions) For any $A \in \texttt{PROD}_\Gamma$ such that $A$ is transitionable into $B$ (with respect to $\Delta$), then $B \in \texttt{PROD}_\Gamma$
    \end{enumerate}
\end{enumerate}
For a tile automata system $\Gamma = (\Sigma,\Lambda,\Pi,\Delta,\tau)$, we say $A \rightarrow^\Gamma_1 B$ for assemblies $A$ and $B$ if $A$ is $\tau$-combinable with some producible assembly to form $B$, if $A$ is transitionable into $B$ (with respect to $\Delta$), if $A$ is $\tau$-breakable into $B$ and some other assembly, or it $A = B$. Intuitively, this means that $A$ may grow into assembly $B$ through one or fewer combinations, transitions, or breaks. We define the relation $\rightarrow^\Gamma$ to be the transitive closure of $\rightarrow^\Gamma_1$, i.e. $A \rightarrow^\Gamma B$ means that $A$ may grow into $B$ through a sequence of combinations, transitions, and/or breaks.
\end{definition}

\begin{definition}[Terminal Assemblies]
A producible assembly $A$ of tile automata system $\Gamma = (\Sigma,\Lambda,\Pi,\Delta,\tau)$ is \emph{terminal} provided that $A$ is not $\tau$-combinable with any producible assembly of $\Gamma$, $A$ is not $\tau$-breakable, and $A$ is not transitionable (with respect to $\Delta$) to any producible assembly of $\Gamma$. We use $\texttt{TERM}_\Gamma \subseteq \texttt{PROD}_\Gamma$ denote the set of producible assemblies of $\Gamma$ which are terminal.
\end{definition}


\section{Simulation Details} \label{sec:simcompute}



In this section, we give more technical details of how $\Gamma = (\Sigma_\Gamma,\Lambda,\Pi,\Delta,\tau)$ is constructed to simulate $\mathcal{A} = (Q,\Sigma_\mathcal{A},\delta,P,\sigma)$.

\emph{State Complexity:} In $\calA$, particles are allowed to know their states and the respective flags of all of their neighbors. Wires are responsible for transmitting signals necessary to coordinate turns between macrotiles. In addition, they must be able to simultaneously transmit signals from their respective clock tiles to neighboring macrotiles, as well as pass any flags or states needed from their respective neighbor to their respective clock tiles. The mechanism with the greatest state complexity(maximal combination of states and flags in $\mathcal{A}$) utilized by tiles takes place during the \emph{state transition} period of a macrotile's turn, which combines the macrotile's flags, $e$ value, $t$ value, and state. Should this transition return  $m = handover_i$, it is similarly responsible for its own flags and the flags of subordinate particle(s). In that case, the results of the state transition in $\Gamma$ that simulates the transition function $\delta$ in $\mathcal{A}$ need to account for up to sixteen flags, two states, one $e$ value, and two $t$ values. The maximal complexity utilized during this procedure occurs within the $handoverPush_i$ signal (detailed in \ref{Signals}), which requires all sixteen flags for all three particles, two states, and two $t$ values to be fully formed, resulting in a maximal tile state complexity of $(\Sigma^{10}\times Q^2 \times e \times t + 1 )$. The constant 1 in the previous expression refers to the $\epsilon$ symbol which is necessarily the $t$ value of the newly contracted subordinate particle.

\emph{Macrotile Mapping:} Each particle $p$ in $\mathcal{A}$ has a unique configuration $C(p) = (\vec{v}, t, e, o, q, (f_0, f_1, ..., f_9))$ that we must be able to represent with our macrotiles. For particle $p$, there exists in $\Gamma$ a macrotile with state $q$ in its central clock tile, flags $(f_0, f_1, ... f_9)$ are represented by the tiles on the extreme ends of the wires extending from the clock bank to each edge, with the value of $f_0$ being represented $o$ faces in the clockwise direction from the northern edge. (A brief note: In our simulation, we define the default orientation $o = 0$ to start labeling pointing north because the grid ($G_\Delta$) on which our macrotiles connect to one another is rotated $30^{\circ}$ from the default orientation of the $(G_\Delta)$ in the geometric Amoebots model. This results in two changes to the neighborhood of a particle on our graph. On our graph, a particle has neighbors in the following directions: (N, NE, SE, S, SW, NW). Tiles on the geometric Amoebots grid have neighbors (E, SE, SW, W, NW, NE), resulting in a replacement of the (E,W) connections with (N,S) in our simulation. Because we couldn't set the default orientation to $o = 0 = E$, we chose another direction in the set of local connections, N.) If the $t$ value of $p \neq \epsilon$, the macrotile corresponding to $p$ is expanded, and has a macrotile in direction $t$ representing its tail. This tail macrotile has the same state $q$ in its central clock tile, and displays on its wire ends whichever subset of five flags corresponds to its relative location. 

\emph{Initial Assemblies and Representation Functions:} Our simulation works as follows with regards to initial assemblies: for every particle $p \in P$, there exists a macrotile in $\Gamma$ that represents $p$ under $R$. This section will show that initial configurations for both systems are isomorphic with respect to $R^*$ and the Simulating Dynamics section ahead will show that $\calA$ follows $\Gamma$ and that $\Gamma$ models $\calA$.


For initial configuration $\sigma$ of $\calA$, we define the initial configuration $\Lambda$ of $\Gamma$ to consist of one large macrotile, itself a connected collection of particle macrotiles, representing $\sigma$ under $R^*$, a number of float tiles equal to at least half of the number of macrotiles in the connected configuration, a number of singleton $\epsilon$ tiles equal to $(6 \times |\mathcal{P}|)$, used to represent empty space in response to query signals, and finally an infinite amount of singleton timing tiles to allow for particle activations. Additional floats above half of the number of positioned particles are fine, more floats simply serve only to relatively speed up perimeter expansion. For every particle $p \in \sigma$, there exists some macrotile in the initial macrotile configuration of $\Gamma$ which maps to an identical configuration $(q \in Q, (f_0, f_1, ..., f_9 \in \Sigma), e \in E, t \in T, o \in O)$ (i.e. it has an identical location and neighborhood, and its state, set of flags, etc. are the same when mapped under $R$).

The macrotile representation function $R$ is defined so that it operates as follows. Given a macrotile $p$, it first checks for the existence of a flag indicating that $p$ is engaged in either a handover or expand move. This is determined by checking the states $p$'s clock and wire tiles. If it is not, $R(p)$ determines the state $q$ from the macrotile mapping to $p$'s central clock tile, values $t$ and $e$ from a subordinate clock tile proximal to the central clock tile, and the flags $(f_0, f_1, ..., f_9)$ from the subordinate clock tile. In the case that particle $p$ is engaged in either a handover or expand move, $R$ instead reads all configuration information from the \textit{configuration tile}, to the west of the clock bank, which holds the particle's previous configuration. This is utilized so that a particle attempting to execute one of these moves will cleanly map to $\alpha$ for the duration of the move, and after the move is completed and the flag removed, will immediately and cleanly map to $\beta$, thus representing it as an atomic move.

\emph{Macrotile Turn:} The simulation of $\calA$ by $\Gamma$ is scaled in both space (by a factor of 100 - the macrotile size) and time. What this means is that, for each atomic asynchronous round of $\calA$, which consists of one particle executing a turn (and in the case of it initiating a handover contract, a series of 4 or 5 turns between it and the subordinate particle depending on whether it is a push or pull), one or two macrotiles of $\Gamma$ go through a series of tile state transitions, additions, and detachments which in total can be considered one atomic operation and represent the round of $\calA$. 

A \emph{turn} is the set of operations a macrotile undergoes in order to simulate a turn for an amoebot particle. In order to capture the asynchronous nature of particle activations within $\mathcal{A}$, we use \emph{timing tiles} to begin a macrotile's turn. Timing tiles float around the simulation and can only interact with an inactive macrotile via a hole in its subordinate clock bank waiting for a timing tile. Upon connection, a timing tile changes the state of the central clock tile from $(q \in Q) \times inactive$ to the same $(q \in Q) \times active$. This state tells the central clock tile to send out $lock$ signals to all of its neighbors in an attempt to begin its move. Should it receive no indication that another particle in its neighborhood is attempting to move, the particle begins to collect the flags of its neighbors indexed to the active particle. In the other case, there must be a neighboring particle attempting to lock down its neighborhood. This will be indicated when the two competing $lock$ signals collide along a wire, either both in one of the tiles' wire, or both will meet on their respective ends of exposed wires from both macrotiles. The resolution of this collision is determined by where the collision takes place. Given macrotiles $\alpha$ and $\beta$, both trying to lock down their neighborhoods with a $lock$ signal somewhere along their shared wire, we will examine possible cases of signal collision. Once a $lock$ signal reaches the end of the wire from which it originated, it is up to the neighboring clock tile to transmit that signal via wire transmission to the neighboring clock tiles. If $lock_\alpha$ were to meet $lock_\beta$, both along $\alpha$'s wire, then the $lock_\beta$ signal would overwrite the $lock_\alpha$ signal, which would  ``bounce" off of the collision and return the central clock tile, returning a negative for $\alpha's$ $lock$ attempt. The inverse would be true should the signals meet while both are in $\beta's$ wire. If both signals meet while both are at the extremis of the wire from which they originated, then both are susceptible to a state transition that would overwrite their own respective signal and allow the neighbor to proceed with their $lock$. If $\alpha's$ $lock$ attempt is overwritten by $\beta's$ $lock$ signal, then upon receipt of $\beta's$ signal, $\alpha$ will send signals the rest of its edges (and by wire transmission, its neighbors) to cancel its own $lock$ order.

Assuming macrotile $\alpha$ may safely continue its turn, it begins to collect the flags indexed to it by its neighbors via query signals sent to its edges. The purpose of these signals is to copy the flag exposed by each of $\alpha$'s neighbors on the exposed wire ends of their shared edges. If $\alpha$ has any empty positions in its neighborhood, the query signals sent down the wires corresponding to these unoccupied locations activate an affinity with the previously mentioned $\epsilon$ tiles, which bind to the exposed wire ends, undergo transitions to indicate that there exists no neighbor at that location, and the signal bounces back from the wire end to the clock. The $\epsilon$ tile changes to a state with no affinity to swarm macrotiles after undergoing its state transition, and so detaches after signalling to the wire end that that location is unoccupied. These signals flow from each wire end back to the clock, where they are successively combined into the subordinate flags/$t$ value signal that represents one half of the tiles that affect the macrotile's simulated amoebot state transition. Once all of this information is combined into a single subordinate clock tile, it undergoes a state transition with the central clock tile, defined in $\Delta$, to produce the new state, flags, and move for the active macrotile. 

If $m = idle$, the macrotile updates its state and flags and then unlocks its neighborhood. After that, it changes its state from $(q \in Q) \times active$ to the same $(q \in Q) \times inactive$, and detaches its timing tile, ending the turn. If $m = Expand_i$, the active macrotile sends a signal to its $i$th edge to allow strength = $\tau$ binding along the exposed wire ends with a float. Only after a float attaches and copies the flags, state, and $t$ value for its new tail does the active tile send out an unlock signal to its neighborhood and ends the turn. Should $m = Contract_i$, assuming the macrotile is expanded in the $i$th direction, it sends a detachment signal to whichever of its ends is detaching and then unlocks its neighborhood after receiving an acknowledgement. Handover contracts are handled in detail in section \ref{Handovers}, but suffice to say they similarly undergo the $lock$ and check procedures, and similarly exchange information between initiating and subordinate macrotiles before unlocking the affected neighborhood and ending the turn. For greater detail on the signals used in this section, refer to Section~\ref{Signals}

\emph{Simulating Dynamics:}
A turn of $\calA$ is simulated in $\Gamma$ via a series of operations which are officially begun once a particle has succeeded in locking its neighborhood. If it receives a timing tile but fails to lock its neighborhood, this does not map to a turn and is the same as if no timing tile had been received (it is an aborted turn). However, at the moment at which it first has officially locked its neighborhood, the series of operations (transitions, tile attachments, and tile detachments) that occur before it unlocks its neighborhood, will map to a single turn in $\calA$.

Consider configurations $A', B'$ of $\mathcal{A}$ such that $A', B' \in \mathcal{C}(P)$ and $A' \rightarrow^\mathcal{A}_* B'$. Any TA system assembly $A \in \texttt{PROD}_\Gamma$ such that under the representation function, $R^*(A) = A'$, must contain a macrotile that maps to each particle in $A'$. Suppose $A' \rightarrow^\mathcal{A} B'$. This means that a single atomic move $m \in M$ from an amoebot is sufficient to change configuration $A'$ to $B'$. Since $\Gamma$ must model $\calA$, it must be the case that there is a pathway for $A$ to transition via the simulation of a turn into $B \in \texttt{PROD}_\Gamma$ such that $R^*(B) = B'$. We will inspect the transitions possible in $A'$ and show how they are modeled in $\Gamma$, so assume $A'$ represents a configuration in between turns and $A$ is an assembly with no macrotiles that are in the middle of a turn (the base case is the initial configuration). Now, let some particle $p$ be selected for the next turn in $A'$. Since $R^*(A) = A'$, we know that whichever macrotile maps to $p$ must represent its configuration, and it also must be able to receive a timing tile. Since no other particle is in the midst of a turn, it will be able to lock its neighborhood and begin simulation of the turn, which will then proceed following the protocol previously described. Interior particles, that is, particles with no exposed edges can only choose $m = idle|\handover_i|contract_i$, assuming the particle has an expanded neighbor or is itself expanded in the $i$th direction. Any macrotile that maps to an interior particle under $R$ is similarly only capable of these moves, since they have no exposed edges to which they could allow floats to attach. $Contract_i$ is only allowed in the case that the particle is expanded.

Any perimeter particle, with exposed wire edges, is capable of any move $m \in M$, assuming the particle's configuration allows for it. This is reflected in the $A$, where perimeter macrotiles are allowed to expand by creating a valid attachment site in the desired direction by changing their exposed wire end in the $i$th direction to a state that has affinity with floats. This, in addition to the moves already available to the interior particles, ensures that any atomic action affected by configuration $A'$ can be represented by a corresponding move in $A$, and the protocol previously described ensures that it can correctly complete, updating all necessary configuration information, before unlocking its neighborhood and thus ending its simulation of the turn. In this way, $\Gamma$ models $\calA$, scaled in space and time, using series of operations that are forced to be atomic are represent all possible atomic moves of $\calA$.

To see that $\calA$ follows $\Gamma$, we refer to the previously described protocol to show that for any $\alpha, \beta \in \texttt{PROD}_\Gamma$ where $\alpha \rightarrow^\Gamma \beta$, then $R^*(\alpha) \rightarrow^\calA_* R^*(\beta)$ and, for idle and contract  moves, until the simulation of the turn completes, $R^*(\alpha) = R^*(\beta)$. For cases where $R^*(\alpha) \rightarrow^\mathcal{A} R^*(\beta)$ via an expansion or handover move, we will explain the steps taken to ensure a clean change from $R^*(\alpha) = R^*(\beta)$ to $R^*(\alpha) \neq R^*(\beta)$. Section \ref{constuctionDefs} defines the \textit{configuration tile}, a tile that is updated with a macrotile's configuration information every turn after a successful move. This tile provides configuration information to the representation function instead of the normal central clock tile and subordinate clock tile that contain a particle's configuration in the case that the active macrotile is engaged in either an \textit{expansion} or $handover_i$ move. $R$ is aware of this because to engage in either of those moves, our construction requires a $handover_i$ or $expansion$ flag to be displayed along the flag tiles of the wire in the $i$th direction.

Only after the simulation of the transition function, for moves $(m = idle, contract_i)$ does $R^*(\alpha) \neq R^*(\beta)$. For moves $(m = expand_i, handover_i)$, $R$ will recognize the intention to expand or handover from the value $m$, and for any particle engaged in these moves, $R$ reads the configuration information from that particle's configuration tile instead of clock tiles, which contains last turn's configuration, until the flags for the move are cleared and the move is complete. In this way, the moves are simulated as atomic operations, and in the case of $handover$s, both atomic and synchronized between the representation of the initiating particle and that of the subordinate particle.
$R^*(\alpha)$ and $R^*(\beta)$ will remain equal until cases of amoebot system ``configuration level changes''. Throughout the simulation of a particle's turn, it may need to send and receive a number of signals from its own wire ends and neighbors' wire ends to ensure that the state transition affected by the central clock tile has all of the information necessary. These signal changes, which are affected by state changes along the wires and subordinate clock tiles, do not change the mapping of configuration $\alpha$, and so $R^*(\alpha) = R^*(\beta)$. However, after a particle undergoes its transition function and changes its state, flags, or makes a move, this is an amoebot system configuration level action. At the moment that these complete and a macrotile begins to idle or contract, the representation of the macrotile changes to $R^*(\beta)$. On the other hand, particles engaged in expansion or handover moves are mapped to $R^*(\alpha)$ until they clear the flag indicating their move from their wires, which takes place after the new states and flags have been distributed and displayed by all particles involved in the move. Every macrotile is capable only of simulating the moves available to the corresponding particle under $R$, since the moves available to a particle depend only on the state, flags, and neighborhood, which must match to be a valid representation under $R^*$.


We further note that float particles which are detached from the rest of the assembly map to empty space and therefore do not impact the represented state of $\Gamma$.
A float functions as a generic ``head" for any particle that wants to expand. Since the amoebots model operates asynchronously and particles are only concerned with local information, they can be simulated with macrotiles similarly concerned only with local information. Since the float tiles do not interact with one another, they effectively float around the configuration which represents amoebot particles, and attach to any site to which a particle which is attempting to expand. Then, upon the next activation, or turn, of that simulated particle, should the particle decide to contract to its new (head) location, the float macrotile stays where it is and the tail macrotile of the particle detaches and becomes another float. Upon detachment, the macrotile queries all of its edges to determine its neighborhood status. Upon recognition that the macrotile has no neighbors, it transitions internally to a default float state, inactive with other tiles with the exception of explicitly labeled valid attachment sites along the perimeter of the swarm. Thus, float tiles simply map to empty space via $R$ unless and until they attach to the assembly representing the swarm and are turned into particle heads or tails.

Since $\Gamma$ models $\calA$ and $\calA$ follows $\Gamma$, $\Gamma$ simulates $\calA$.

\subsection{Signals Between macrotiles}
The following section details all of the signals utilized by macrotiles to accurately simulate amoebot behavior. With the exceptions of the \emph{Start Turn Signal} and \emph{End Turn Signal}, which are propagated via state transitions within the clock, these signals are propagated via wire transmission from one central clock bank of a macrotile to a neighboring clock bank via wire transmission. With the exception of any signals pertaining to $handover_i$, these signals do not have the capability to change the state, flags, or $t$ value of any particle macrotile from which they did not originate. 

\subsection{List of Signals} \label{Signals}
\begin{enumerate}
    \item{\emph{Lock $Signal_i$}}: To prevent conflicts, we have macrotiles engaged in their turn send signals to all particles that may be affected by the turn that ``lock" the tiles into their current configuration. A locked macrotile will not activate even if it receives a timing tile until it receives an unlock signal from the same direction that sent the lock signal. If two lock signals from different macrotiles collide along a shared wire, the resolution depends on where they collide. If neighboring macrotiles $a$ and $b$ both send lock signals to each other, and these signals meet along $a$'s half of the shared wire, then the signal from $b$ will overwrite the lock signal from $a$. The opposite is true should they meet inside $b$'s wire. If both signals are in the respective extreme tiles of their own wires, then both are subject to a state transition that will allow either $a$'s signal to propagate into $b$'s wire, allowing $a$ to expand, or $b$'s signal into $a$'s wire, in which case $b$ would be safe to expand. Should an expanded neighbor of a macrotile receive a $lock$ signal, it further propagates this signal to its other half, to prevent it from attempting to move and conflict with the active particle.
    
    Let us represent the lock signal as state $L$ in $\Gamma$, which both clock and wire tiles utilize in addition to the states in $\Sigma$ for the system they are simulating. In order to enable these tiles to maintain both their respective states in $\Sigma$ and any signals they need to pass, we cross the set of states representing signals utilized in our construction of $\Gamma$ by the set of signals and states $(\Sigma, Q)$ in $\mathcal{A}$. When a clock tile undergoes a transition with a wire tile to indicate that the wire should start propagating a lock signal, the wire tile adjacent to the subordinate clock tile changes states from its previous state $f \in \Sigma$ to $Lf$, a cross of signal $L$ and any potential flag $f \in \Sigma$. This signal is propagated down the wire until it hits the exposed flag tile at the end of the wire. If the exposed flag tile at the end of a wire is not connected to a neighbor, the state displayed $Lf$ has at least $\tau$ strength affinity with $\epsilon$ tiles, enabling connection. If an epsilon tile attaches to an exposed wire end displaying $Lf$, both tiles undergo a directed double state transition which indicates to the flag tile that there is no neighbor at that location, and simultaneously transitions the $\epsilon$ tile to a state without affinity to anything else, leaving it inert and detaching it from the flag tile at the perimeter of the swarm. If a flag tile of macrotile $a$ displaying a lock signal is connected to a flag tile of neighboring macrotile $b$, it waits as is until it receives a confirmation signal $L'f'$ from $b$'s respective flag tile. Upon receiving the $Lf$ signal from $a$'s flag tile, $b$'s flag tile starts to propagate a cross of its flag $f' \in \Sigma$ and a state indicating that a neighbor has sent a lock signal, $Lnf'$. When this signal propagates all the way to $b$'s clock bank, the innermost wire tile undergoes a double transition with its respective subordinate clock tile, leaving the wire tile in state $Ln'f'$, and the clock tile in its previous state crossed with the neighbor lock signal state $Ln$. The clock tile that received the $Ln$ signal further propagates it to the rest of $b$'s clock, so that $b$ will not respond to timing tiles should they attach for the duration of the $a$'s move. The new state $Ln'$ crossed with the wire's existing state $f'$ indicates that the clock bank of $b$ has received the lock signal, and is propagated from $b$'s clock bank back to the $b$'s flag tile facing $a$. Once $a$'s flag tile observes the response $Ln'f'$ from $b$'s flag tile, it propagates the confirmation signal $L'f$ back along the wire to $a$'s clock bank. The central clock tile in $a$ maintains a state indicating how many responses to its lock signal it has received. Once $a$ has received six or ten responses (depending on whether it is contracted or expanded), it continues its move by sending out a $query$ signal to all of its wires, to find out the flags displayed to it by its neighbors.

    \item{\emph{Progressive Lock $Signal_i$}} This signal is functionally identical to the preceding $lock Signal_i$, except that when macrotile $a$ sends this signal to macrotile $b$, $b$ further propagates $lock$ signals to its own respective neighborhood. $b$'s neighborhood remains locked down until it receives the $Unlock Signal_i$ from $b$, itself contingent upon receiving the $Unlock Signal_i$ from $a$. This is utilized in $handoverPull_i$ moves, which require the initiating particle to lock down both its own and the subordinate particle's neighborhoods. 
        
    \item{\emph{Unlock $Signal_i$}}: The other half of the Lock Signal, this tells a neighborhood that they may begin their turns upon receiving a timing tile. This does not require a response from any macrotile receiving it, as any macrotile unlocking its neighborhood is done with its turn and will not attempt to move until its next turn.

    \item{\emph{Copy Signal}}:  When a float attaches to the swarm, it attaches via a single bond to a specific tile that is actively trying to expand. In order for the float to become the head of the particle to which it is attaching, it needs to display the proper flags and state. In the cases where an active macrotile is attempting to expand along any port other than its $0$th port, the copy signal consists of the active macrotile's state $(q \in Q)$ and the flags pertinent to the new head $(f_i, f_{i+1}, ..., f_{i+4})$. We know which flags are important to the new head because the new head will always be in the $i$th direction, and we only need that direction's flag as well as the next four, as the particles consume two of their potential edges via their connection. This heuristic for which flags to send (i.e. sending the $i$th flag and the next four) to the new head breaks down if a particle is attempting to expand along its $0$th port. In this case, the copy signal consists of the active macrotile's state $(q \in Q)$ and flags $(f_0, f_1, f_2, f_8, f_9)$. In this case we make the edge of the head corresponding to the $0$th port in the expanding macrotile the new $0$th port, since we can't start labeling available edges along the shared edge of an expanded macrotile. These states and flags are propagated via wire transmission and stored in the new central clock tile, wire extremities, and subordinate clock tiles, respectively.

    
    \item{\emph{Attachment Signal}}: Given macrotile $a$ and an adjacent unoccupied location $a'$ on $G_\Delta$, this signal is sent from the clock bank of macrotile $a$ to the wire corresponding to the edge shared between $a$ and $a'$. Upon reaching the exposed end of the wire, this signal compels the flag tile to change states to a state with $\tau$-strength affinity for a float tile's exposed wire ends. $a$ does not send the $UnlockSignal$ signal corresponding to the $LockSignal$ that preceded its expansion until the float assumes its flags and state, facilitated by a $Copy Signal$. The $Acknowledge Signal$ following this $Copy Signal$ causes a state change in the flag tile of the newly connected wire which is propagated back to $a$'s clock bank, allowing it to safely unlock its neighborhood and end its turn.
    
    \item{\emph{Detachment Signal}}: This signal is sent to the macrotile of an expanded particle that is not the macrotile occupying the $G_\Delta$ location to which the particle is contracting. It tells the macrotile to switch its internal state and flags to the default state and flags, which will necessarily detach it from the swarm, as the default flag has no affinity with the swarm. After detaching, the detached macrotile will query its neighborhood and transition its state and flags to those of a float after receiving an $\epsilon$ response from all of its edges. 
    
    \item{\emph{$HandoverConfirm_i$}}: This signal is sent from an initiating macrotile in the $i$th direction to check to ensure that the $i$th neighbor matches the orientation necessary to complete the move (i.e. a macrotile isn't trying to $handoverPull_i$ with a neighbor that is already expanded, as that would violate the constraint of a particle occupying at most two locations on $G_\Delta$). This requires an acknowledgement signal containing the subordinate macrotile's $t$ value to ensure validity. After receipt of a valid response from the subordinate macrotile, the signal sent next by the initiating supertile depends on whether the initiating particle is executing a $handoverPush_i$ or a $handoverPull_i$. This signal persists at the threshold of the wire shared between the subordinate particle and the particle changing hands until all necessary state and flag information has been passed successfully, ensuring the tiles engaged in this move are mapped to their previous configurations. This ensures a clean mapping from one configuration for a macrotile to another under $R$, which will take place instantaneously after the $handover_i$ flag is cleared from the macrotiles' respective wires.
    
    \item{\emph{$HandoverPull_i$}}: Consider an expanded macrotile $a$ executing a $HandoverPull_i$ move with a contracted macrotile neighboring its tail $b$. After $b$ sends a valid $Acknowledgement Signal$ containing its $t$ value and $a$ decides to continue its turn, $a$ sends a $New Head Signal$ to the macrotile of which it is ceding control. $a$ then sends the ten flag values $(f'_0, f'_2, ... f'_9) \in \Sigma$ and new state $q' \in Q$ combined into a single signal to $b$. Upon receiving this signal, $b$ sends out a query signal to each of its edges to determine which of its neighbors is its new head. After discovering its new $t$ value, $b$ updates its $t$ value, state, and sends its new flags to their respective ports. $b$ then sends an $AcknowledgementSignal$ to $a$ indicating that it has assumed control of the macrotile that changed hands, and $a$ sends out an $Unlock$ signal, which is further propagated out to all of $a$ and $b$'s neighborhoods. 
    
    \item{\emph{$HandoverPush_i$}}: Given a contracted macrotile $a$ executing a $HandoverPush_i$ move with an expanded neighboring macrotile $b$, assuming both are the proper configurations, $a$ sends to its newly acquired head the following bundled into a single $HandoverPush_i$ signal: its own flags $(f_i, f_{i+1}, ... f_{i+4} \in \Sigma)$, its state $(q \in Q)$, and the new $t$ value for the newly acquired head; new flags $(f'_0, f'_1, ... f'_5 \in \Sigma)$, a new state $(q' \in Q)$, and $t = \epsilon$ for the subordinate macrotile $b$. The subordinate macrotile strips the information pertinent to itself and propagates the rest of the signal further on to its other half. After the other half sends back an $Acknowledgment Signal$ indicating that it has updated, $a$ sends out an unlock signal which is further propagated out to all of $a$ and $b$'s neighborhoods. 
    
    \item{\emph{$New Head Signal$}}: In $handoverPull$ moves, part of an expanded particle is ceded to an inactive, contracted particle, thereby simultaneously contracting the initiating particle and expanding the subordinate particle. This signal is sent to the macrotile segment of an expanded particle that is changing owners, telling it to display a special flag on all ports indicating that it is a newly acquired ``head" particle, allowing the subordinate macrotile to identify which of its neighbors now serves as its head.
    
    \item{\emph{Query Signal}}: This signal is sent by the clock tiles to the ends of their wires to collect the flags displayed by each of the originating macrotile's neighbors. Naturally, its domain is $\Sigma \cup \{\epsilon\}$, with the empty flag returning in the case that that edge has no neighbor via the usual mechanism of singleton $\epsilon$ tiles.
    
    \item{\emph{Acknowledgement Signal}}: Used by particles to respond to requests for information not displayed on the flags, such as state or $t$ value. Additionally used as a confirmation signal with certain $handover$ operations. 
    
    \item{\emph{Start Turn Signal}}: Any macrotile that is inactive and is not locked down by a neighbor has an empty tile in its clock bank waiting for the attachment of a timing tile. Once this timing tile attaches, it uses state transitions to ``activate" the central clock tile, which in turn propagates this signal down all of its wires so that it can read its neighbors' flags indexed to it.
 
    \item{\emph{End Turn Signal}}: After a macrotile unlocks its neighborhood after a move, it sends a signal from the central clock tile to the timing tile to detach, deactivating the particle until next time a timing tile diffuses into its slot and it is not locked down. This constitutes the \textit{End Turn signal}.
    
\end{enumerate}

\section{Communication Protocol}
This section explores an example turn where contracted macrotile $a$ executes a $handoverPush$ with the tail of its expanded neighbor $b$. We will list all signals that will be sent and received by all macrotiles involved, in order. We chose to use a $handover$ move as our example because they require the greatest number of signals as well as the greatest signal complexity of any moves available to particles. For the purposes of this illustration, assume particles start in an inactive state, with their proper flags displayed on their respective flag tiles. Figure \ref{fig:Example_turn} shows some condensed snapshots from the turn explored in the following section.

\subsection{Example turn}
\begin{enumerate}

\item Timing tile attaches to $a$'s clock bank and precipitates a transition, sending the \textit{Start Turn Signal} to $a$'s central clock tile and changing it from $(q \times inactive)$ to $(q \times active)$.
\item The central clock tile of $a$ checks its $t$ value held in the state of the tile to its immediate west to determine how many responses it should expect and then sends out the \textit{lock signal} to all of its wires. $a$ is contracted, and so its $t$ value = $\epsilon$, meaning that it should expect six responses from its wires. The central clock tile maintains both $q$ and a response counter for the duration of this signal.
\item After $a$ has received six responses indicating that none of its neighborhood has sent a competing \textit{lock signal}, it sends out a \textit{query signal} to gather the flags indexed to it by $a$'s neighbors. This information is necessary for $a$ to affect the transition function $\delta$ in $\mathcal{A}$.
\item When the \textit{query signal} reaches the end of each of $a$'s wires, it causes the flag tiles at the extreme end to change to a state receptive to reading the state displayed on its neighbor's flag tile. If a flag tile has no neighbor, the \textit{query signal} state also enables connection with $\epsilon$ tiles, which will connect and undergo a transition to indicate that the wire corresponds to an unoccupied edge.
\item After six flags or $\epsilon$'s are propagated back to the tile to the west of $a$'s clock bank, the central clock tile undergoes a double transition with the subordinate clock tile to its immediate west. $a$'s state contains the macrotile's state, $q$, and the subordinate tile to the west's state contains the combined six flags of $a$'s wires as well as the $t$ value.
\item After the state transition effecting both the central clock tile and the subordinate clock tile to its west, the central clock tile contains the new state, $q \in Q$, as well as the move $m = handoverPush_i$, where $i$ is the port in the direction of $b$'s tail. The tile to the west contains the new flags for $a$, $(f_0, f_1, ..., f_9) \in \Sigma^{10}$, particle $a$'s $E$ value = $i$, the new flags for $b$, $(f'_0, f'_1, ..., f'_5) \in \Sigma^6$, $b$'s new state $q' \in Q$, and the new $t$ value for $b = \epsilon$.
\item Because $m = handoverPush_i$, $a$ needs to check its neighbor in the $i$th direction to ensure that they have the proper configuration to enable to move. It sends a $handoverConfirm_i$ signal to the neighbor in its $i$th direction and waits for an \textit{Acknowledgement signal} in response containing $b$'s $t$ value.
\item $b$'s tail responds by sending $b$'s $t$ value in an \textit{Acknowledgement Signal} to $a$.
\item Since $b$'s $t$ value $\neq \epsilon$, b is a valid neighbor with which $a$ can execute the $handoverPush_i$ movement. $a$ continues by sending a $handoverPush_i$ signal along its $i$th port, to $b$'s tail. This signal contains five new flags $(f_i, f_{i+1}, ... f_{i+4}) \in \Sigma$, as well as a copy of $a$'s state $q$ and the new $t$ value for $b$'s tail, soon to become $a$'s head. It additionally contains six flags $(f'_0, f'_1, ..., f'_5)$ as well as the new $t$ value = $\epsilon$ for $b$'s head.
\item After receiving the $handoverPush_i$ signal from $a$, at least one of $b$'s wires will display a flag indicating that it is engaged with a handover move, additionally, $b$'s tail updates its state, $t$ value and flags. It further propagates the $handoverPush_i$ signal to $b$'s head. $b$'s tail will continue to map to $b$ until the final signal ending the move clears the any flags indicating to the representation function that the macrotile displaying that flag is undergoing an expansion or $\handover$ move. 
\item Once $b$'s head receives the $handoverPush_i$ signal from its tail, it updates its state, flags, and $t$ value as well. After doing so, it sends an \textit{Acknowledgement signal} to $b$'s tail, indicating that it has updated its state, flags, and $t$ value.
\item $b$'s tail propagates $b$'s head's \textit{Acknowledgement signal} back to $a$. This acknowledgement signal clears any flags indicating a macrotile's partaking in an expansion or $\handover$ move, and so after it passes through what was formerly $b$'s head and tail, those tiles respectively map to $b$ and $a$'s head.
\item Upon receiving the \textit{Acknowledgment signal} indicating that the move has been completed, $a$ sends out an unlock signal to its neighborhood. 
\item The newly expanded macrotile $a$ detaches the its timing tile by the central clock tile transitioning the timing tile from an \textit{active} state to an \textit{inactive} state, which has no affinities.

\begin{figure}[h]
\centering
\includegraphics[width=\textwidth]{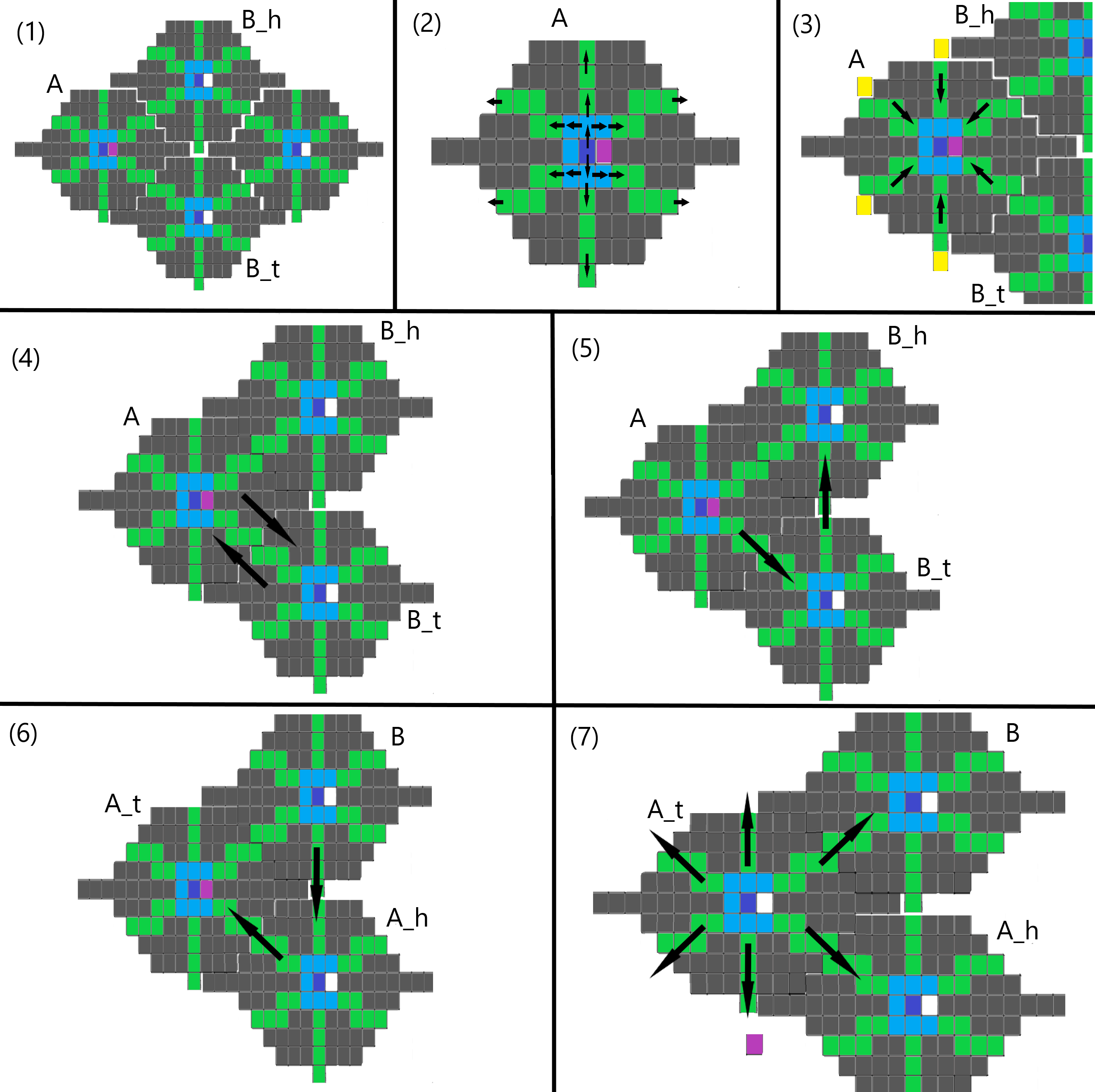}
\caption{(1)The initial configuration of a subset of a valid swarm. Note that only A has a purple timing tile to the right of its Central Clock tile. (2) The flow of the $Lock Signal_i$ sent out from A's CC tile. (3) The responses from A's wires flow back to the CC tile. Unoccupied edges allow for the attachment of yellow epsilon tiles, which indicate to a given wire that it has no macrotile neighbor. A has no neighbors attempting their own move, and is safe to continue its move. (4) A is attempting to execute a handover, and so much check to ensure that the subordinate particle it intends to use is in the proper configuration. It sends out a $handoverConfirm_i$ signal, to which B responds with an $Acknowledgement$ signal containing B's $t$ value. (5) Since B is in the proper configuration to enable a $handoverPush_i$, A continues by sending a $handoverPush_i$ signal to the subordinate macrotile it wants to take over. The subordinate macrotile $B_t$ strips the flags, state, and new $t$ value pertinent to itself and propagates the remaining new flags, state and $t$ value to its other half $B_h$. (6) $B_h$ alters its state and flags in response to the signal it has just received and sends back an acknowledgement signal to its tail. As this signal travels through the shared wires between $B_h$ and $B_t$, it clears the $handover$ flags which indicate to the representation function that it should check those macrotiles' configuration tile for information instead of flags and the CC tile for the state. The clearing of these flags is the event that precipitates the transition of the representation of the macrotile displaying them, so that $R^*(\alpha) \neq R^*(\beta) $, where $\alpha$ and $\beta$ are the macrotile before and after this step. (7) Finally, $A_t$ receives the acknowledgement signal propagated from what now maps to $A_h$, so it sends out a final $Unlock$ signal to all of its neighborhood and A ends the turn by the CC undergoing a transition with the timing tile, rendering inert and without affinities, thereby detaching it. }
\label{fig:Example_turn}
\end{figure}

\end{enumerate}

\end{document}